\documentclass[12pt,a4paper]{article}
\usepackage[latin1]{inputenc}
\usepackage{amsmath}
\usepackage{amsfonts}
\usepackage{amssymb}
\usepackage{bbm}
\begin{document}

\begin{titlepage}
\begin{center}
{\LARGE Higgs picture of the QCD-vacuum}

\vspace{2cm}
C. Wetterich\footnote{e-mail: C.Wetterich@thphys.uni-heidelberg.de}\\
\bigskip
Institut  f\"ur Theoretische Physik\\
Universit\"at Heidelberg\\
Philosophenweg 16, D-69120 Heidelberg\\
\vspace{3cm}
\today
\end{center}

\begin{abstract}
The functional integral for QCD is reformulated by introducing
explicitly an integration over the fluctuations of composite
quark-antiquark bound states. Chiral symmetry breaking by the color
singlet scalar field induces masses for the fermions. Our formulation
with scalar fluctuations may be useful for lattice gauge theories by
modifying the spectrum of the Dirac operator in the vacuum and
permitting a simple connection to chiral perturbation theory. We
propose that a ``condensate'' of quark-antiquark bound states in the
color octet channel generates masses of the gluons by the Higgs
mechanism. A simple effective action for quarks, gluons and
(composite) scalars yields a surprisingly good description of the
charges, masses and interactions of all low mass physical excitations
- baryons, pseudoscalars and vector mesons. Dressed quarks appear as
baryons and dressed gluons as vector mesons.
\end{abstract}
\end{titlepage}


\section{Introduction}
\label{introduction}

An analytic description of the vacuum in QCD remains a central goal in
quantum field theory. We have witnessed convincing progress of
numerical simulations of QCD on a lattice. Still, even an approximate
analytic understanding would be a highly valuable complement.
Computations of strong cross sections and decay rates are very hard in
lattice QCD. Simulations have a long way to go before the properties
of nuclei can be explained. Furthermore, a huge present and future
experimental program tries to gather information about the QCD phase
transition and the phase diagram as function of baryon density and
temperature. Simulations for high baryon density are notoriously
difficult and an analytical understanding would be very helpful.

A key issue for the simplicity and success of an analytical
description is an efficient description of the relevant degrees of
freedom. In QCD the relevant degrees of freedom depend on the momentum
scale. Processes involving high momenta (more precisely high
virtuality) are well described by quarks and gluons. Perturbation
theory based on the microscopic action for quarks and gluons with a
small gauge coupling yields reliable results. In contrast, the
relevant degrees of freedom at long distances  or small momenta are
mesons and baryons. An efficient description of the vacuum should
therefore involve degrees of freedom for the low mass
mesons. Typically, the lowest mass excitations in the QCD vacuum
should comprise the pseudoscalar meson octet and singlet $(\eta')$,
the vector meson octet and the baryon octet in the respective
channels. We observe that this spectrum of real QCD differs strongly
from pure QCD (gluodynamics) without quarks where the low excitations
consist of glueballs. This simple fact suggests that an efficient
analytic description of real QCD at low momentum differs substantially
from gluodynamics. As the momentum scale is lowered an analytic
description of real QCD should effectively switch from gluons and
quarks to mesons and baryons.

The most prominent long distance degrees of freedom are scalar
quark-antiquark $(\bar q q)$ bound states. The expectation  value of
such a composite scalar field induces spontaneous chiral symmetry
breaking. The excitations of this scalar field describe then the
associated (pseudo-) Goldstone bosons $\pi,\,K,\,\eta$ as well as the
$\eta'$. Our first task for an analytic description will therefore be
to supplement quarks and gluons by the degrees of freedom for
composite scalar (and pseudoscalar) meson fields. In the first part of
this work (sect. 2) we will reformulate the functional integral for
QCD in an exactly equivalent form which comprises an integration over
explicit scalar degrees  of freedom. These scalars are of minor
importance at high momenta but become crucial for the properties of
the vacuum. Besides the analytical  advantage of a simple connection
to chiral perturbation theory this reformulation may also offer
important benefits for lattice simulations. The spectrum of the Dirac
operator  acquires a mass gap even in the chiral limit and the contact
to chiral perturbation theory at long distances should become
straightforward.

Concerning the QCD phase transition the task for an analytical
description becomes even more involved: the formalism should now
describe {\it simultaneously} quarks, gluons, mesons and baryons. Above
the critical temperature a quark gluon plasma is a valid approximation
and explicit mesons or baryons play no important role. In the hadron
gas below the critical temperature the mesons (and baryons) dominate
whereas quarks and gluons become irrelevant. A simple analytical
description has to capture all these degrees of freedom and provide
for a mechanism explaining why their relative importance changes
abruptly as a function of temperature.

Associating spontaneous chiral symmetry breaking with the phase
transition yields a simple explanation why the  quarks
disappear from the relevant spectrum below the critical
temperature. $T_c$: the fermions get massive for $T<T_c$.
What is needed is a similar mechanism for the gluons. For
the electroweak phase transition the effective generation of a mass for
the $W$- and $Z$-bosons is well known. As the universe cools down
below the critical temperature both the gauge bosons and the fermions
acquire a mass due to the Higgs mechanism associated to the
``spontaneous breaking'' of the electroweak gauge symmetry. The second
part of this work will review a similar ``Higgs mechanism'' for QCD,
namely the ``spontaneous breaking of color'' by the expectation value
of composite $\bar q q$-scalars in the color-octet channel
\cite{CWSSB}.

Strictly speaking, local symmetries cannot be broken spontaneously
in the vacuum. This has led to the realisation that the
confinement and the Higgs description are not necessarily associated to
mutually exclusive phases. They may only be different
facets \cite{Banks} of one and same
physical state\footnote{The complementarity between the Higgs and
 confinement description has been considered earlier for toy models
 with fundamental colored scalar fields \cite{H}.} We stress that this
 observation
is not only of formal importance. For example, the high temperature
phase transition of electroweak interactions with a small Higgs scalar
mass ends for a larger scalar mass in a critical endpoint. Beyond
this endpoint the phase transition is replaced by an analytical crossover
\cite{ReWe}. In this region -- which is relevant for a realistic Higgs
mass in the standard model -- a Higgs and a confinement description
can be used simultaneously. Our picture of the QCD vacuum resembles in many
aspects the ``strongly coupled electroweak theory'' at high
temperature\footnote{Without a direct connection to the
standard model the $SU(2)$-Yang-Mills theory with strong
gauge coupling and fundamental scalar has been first simulated
on the lattice in \cite{SU2L}.}. We will present a Higgs description
of QCD as a complementary picture to the usual confinement
picture. Both the Higgs description and the confinement picture are
considered as valid descriptions of one and the same physical
properties of QCD.  A valid Higgs picture has to be consistent with
the well established results of the confinement picture and of lattice
QCD.

With respect to color the composite scalar $\bar q q$-bilinears
transform as singlets and octets, each in the $(\bar 3,3)$-representation
of the $SU(3)_L \times SU(3)_R$ chiral flavor symmetry. We will
explore the hypothesis that the scalar octets induce ``spontaneous
color symmetry breaking'' while a ``physical'' global $SU(3)$ symmetry
is preserved. This global symmetry can be used to classify the
spectrum of excitations according to the ``eightfold way''. We find
that all gluons acquire a mass by the ``Higgs mechanism'' and belong to
an octet of the physical $SU(3)$-symmetry. Our picture provides an
effective infrared cutoff for real QCD by mass generation. (The cutoff
for gluodynamics is expected to be different!)

The Higgs mechanism also gives integer electric charge and strangeness
to all physical particles according to their $SU(3)$-transformation properties.
 Furthermore, the expectation value of the
quark-antiquark color octet breaks the global chiral symmetry. In
consequence the fermions become massive and the spectrum contains
light pions and kaons as pseudo-Goldstone bosons. In the limit of
equal masses for the three light quarks the global vector-like
SU(3)-symmetry of the ``eightfold way'' becomes exact. The nine quarks
transform as an octet and a singlet.  We identify the octet with the
low mass baryons -- this is quark-baryon duality. 
The singlet is associated with the $\Lambda$ (1405) baryon and has
parity opposite to the nucleons.
Similarly, the eight
gluons carry the quantum numbers of the light vector mesons
$\rho,K^*,\omega$.  The identification of massive gluons with the
vector mesons is called gluon-meson duality.

We propose that the main nonperturbative new ingredient for the
effective action
of low-momentum QCD consists of
  scalar fields representing  quark-antiquark bound states.
Once these composite operators are treated on the
same footing as the quark and gluon fields, the description of
propagators and vertices  becomes again
very simple. We will investigate the phenomenological  consequences of
an effective action that adds to the usual gluon and quark (gauge covariant)
kinetic terms the corresponding kinetic terms for the scalars with
quantum numbers of the $\bar q q$-composites. Furthermore, this is
supplemented by a scalar potential and a Yukawa interaction between
quarks and scalars. It is very remarkable that such a simple effective
action can account for the quantum numbers, the masses and the
interactions of the pseudoscalar octet and the $\eta'$ mesons, the vector
meson octet and the baryon octet! Once the free parameters of this
effective action are fixed by matching observation, several nontrivial
properties of hadrons can be ``predicted'' without further unknown
parameters. These findings suggest that the
long-sought  dual description of long-distance
strong interactions can be realized by the addition of fields for composites.
The situation would then be quite
 similar to the asymptotically free nonlinear sigma model
in two dimensions where the addition of the composite ``radial excitation''
provides for  a simple dual description of the low momentum behavior
\cite{1}. 

Spontaneous breaking of color has also been proposed \cite{4} for
situations with a very high baryon density, as perhaps in the interior
of neutron stars. In this proposal a condensation of diquark operators
is responsible for color superconductivity and spontaneous breaking
of baryon number. In particular, the suggestion of
color-flavor locking \cite{CF} offers analogies to our description
of the vacuum, even though different physical situations are
described (va\-cuum vs. high density state) and the pattern of spontaneous
color-symmetry breaking is distinct (quark-antiquark vs. quark-quark
condensate; conserved vs. broken baryon number). This analogy
may be an important key for the understanding of possible phase
transitions to a high density phase of QCD.

In the second part of this work we first present in sect. 3 our
proposal for a simple effective action for QCD. Sect. 4 describes the
``Higgs picture'' with a non-vanishing expectation value  of the
scalar octet field. The gauge invariant description in terms of
nonlinear fields  is introduced in sect. 5. Sect. 6 discusses the
role of a hidden local symmetry in the nonlinear description which
will permit a direct connection to ideas of ``vector dominance''. The
elctromagnetic interactions can be used as an efficient probe of our
picture since no new free parameters are introduced (sect. 7). In
sect. 8 we turn to the interactions of the vector mesons and the
decay $\rho\to 2\pi$. Here we also make direct contact with the
description  of vector mesons as gauge invariant  $\bar q\gamma^\mu q$
bound states in sect. 2. Sect. 9 discusses the interactions of the
pseudoscalar mesons. We will see that the strength of the vector and
axialvector couplings of the nucleons as well as the pion interactions
beyond leading order chiral perturbation are successfully accounted
for by our simple effective action. Sect. 10 finally presents
conclusions and discussion and makes a simple proposal for the QCD
phase diagram.

\section{Functional integral with composite fields}
\label{functionalintegral}
Our starting point is the partition function for QCD
\begin{equation}\label{A1}
Z=\int{\cal D}\psi{\cal D} A~e^{-S}
\end{equation}
where $S$ is the gauge invariant classical action for quarks and
gluons. Here the quarks are described by Grassmann variables obeying
$\left\{\psi(x),~\psi(y)\right\}=0 ~~,
~~\left\{\psi(q),~\psi(q')\right\}=0$ in position and momentum space,
respectively. The difficult part in the functional integral (\ref{A1})
is the functional measure $\int{\cal D}\psi {\cal D} A$ which includes
the regularization , for example on a lattice, or the gauge fixing and
ghost parts for a continuum formulation. We do not need a
specification here and only assume that the functional measure preserves
the gauge symmetry.

It is well known that local gauge symmetries cannot be spontaneously
broken (in a strict sense). Therefore only gauge invariant quantities
can have nonzero expectation values. A prominent example concerns the
correlation functions for scalars and pseudoscalars which are
contained in
\begin{equation}\label{A2}
G_s(x,y)=-\langle \left(\bar{\psi}_L(x)\psi_R(x)\right)
\left(\bar{\psi}_R(y)\psi_L(y)\right)\rangle.
\end{equation}
Here $(\bar{\psi}_R\psi_L)=(\bar{\psi}\frac{1+\gamma^5}{2}\psi)$ and
brackets denote contractions of spinor and color indices, while we
have not displayed the (open) flavor indices. For the example of the
pion channel the two point function decays for large $|x-y|$ as
$G_s\sim\exp\left(-m_\pi|x-y|\right)$ and this is the way how the pion
mass is measured, for example on the lattice. Similarly, the
$\rho$-meson mass can be extracted from the gauge invariant
correlation function in the vector channel
\begin{equation}\label{A3}
G^{\mu\nu}_V(x,y)=\langle\left(\bar{\psi}(x)\gamma^\mu
\psi(x)\right)\left(\bar{\psi}(y)\gamma^\nu\psi(y)\right)\rangle.
\end{equation}

A convenient way for the computation of correlation functions for
gauge invariant quark-antiquark bilinears is the introduction of gauge
invariant sources
\begin{equation}\label{A4}
S_j=-\int d^4x\Big\{(\bar{\psi}_L(x)j^\dagger_s(x)\psi_R(x)+h.c.)
+\bar{\psi}(x)\gamma^\mu j_{V_\mu}(x)\psi(x)\Big\}.
\end{equation}
Here $j_s$ is a complex $N_f \times N_f$ matrix and $j_V$ denotes a
hermitean $N_f \times N_f$ matrix, with $N_f$ the number of quark
flavors. We will concentrate here on the three light flavors of quarks
$(N_f=3)$ while considering the two flavor case as a pedagogical
example below. Then the physical values of the scalar sources are given by
the current quark masses
\begin{equation}\label{A5}
j_s=j^\dagger_s=m_q= \left(
\begin{array}{lcr}
m_u&&\\ &m_d&\\ &&m_s
\end{array}\right)
\end{equation}
while the vector source vanishes $j_V=0$. Adding $S_j$ to $S$ the
partition function $Z[j]$ (\ref{A1}) becomes now a functional of the
sources.

As an example, we may write the expectation value of the scalar
quark-antiquark bilinear as ($a,b$ are flavor indices while color and
spinor indices are contracted)
\begin{equation}\label{A5a}
\langle\bar{\psi}_{Lb}(x)\psi_{Ra}(x)\rangle=
\sigma_{ab}(x)=Z^{-1}\frac{\delta Z}{\delta j^*_{ab}(x)}.
\end{equation}
Similarly, the (unconnected) two point function obtains from the
matrix of the 
second functional variations\footnote{In flavor space $(G_s)_{ab,cd}$
  is a $N^2_f\times N^2_f$ matrix, with rows and columns labeled by
  flavor index pairs $(ab)$ and $(cd)$.}
\begin{eqnarray}\label{A6}
G_s^{(u)}(x,y)&=&-\langle\left(\bar{\psi}_R(x)\psi_L(x)\right)
\left(\bar{\psi}_L(y)\psi_R(y)\right)\rangle\nonumber\\
&=&Z^{-1}\frac{\delta^2Z}{\delta j(x)\delta j^*(y)}.
\end{eqnarray}
As usual, the expectation value or the connected Green's function can
be derived from
\begin{equation}\label{A7}
W[j]=\ln Z[j],
\end{equation}
as
\begin{equation}\label{A8}
\sigma_{ab}(x)=\frac{\delta W}{\delta j^*_{ab}(x)}
\end{equation}
or
\begin{equation}\label{A9}
G_s(x,y)=G^{(u)}_s(x,y)-\sigma^*(x)\sigma(y)=
\frac{\delta^2 W}{\delta j(x)\delta j^*(y)}.
\end{equation}
Finally, the effective action $\Gamma$ is a functional of the
``classical field'' $\sigma$
\begin{equation}\label{A10}
\Gamma[\sigma]=-W[j]+\int
d^4x~tr~\big(j^\dagger(x)\sigma(x)+\sigma^\dagger
(x)j(x)\big)
\end{equation}
where $\sigma$ corresponds to a given source $j$ and $ j=j[\sigma]$ is
computed by inversion of
\begin{equation}\label{A11}
\sigma[j]=\frac{\delta W}{\delta j^*}.
\end{equation}
The field equation reads
\begin{equation}\label{A12}
\frac{\delta\Gamma}{\delta\sigma_{ab}(x)}=j^*_{ab}(x).
\end{equation}
Furthermore, the second functional variation of $\Gamma$ with respect
to $\sigma$ equals the inverse connected two point function, i.e. the
inverse of the second functional variation of $W$ with respect to $j$
\begin{equation}\label{A13}
\Gamma^{(2)}W^{(2)}=\mathbbm{1}.
\end{equation}
(Note that for this matrix notation all internal and space or momentum
labels of $\sigma$ are collected into a vector.) Of course, this
setting can be generalized to other gauge invariant composite
operators, like vector fields, in a straightforward way.

Let us concentrate here on the scalar sector and, for simplicity of
the demonstration, on $N_f=2$. For $x$-independent values
$\sigma(x)=\sigma$ one has $\Gamma=\int d^4xU(\sigma)$ where the
effective potential $U$ is now a simple function of the complex $2x2$
matrices $\sigma$. By construction $U(\sigma)$ is invariant under the
chiral flavor rotations $SU(N_f)_L\times SU(N_f)_R$. If we expand
$U(\sigma)$ in powers of $\sigma$ only invariants can appear
\begin{eqnarray}\label{A14}
U(\sigma)&=&m^2~ tr(\sigma^\dagger\sigma)-\frac{\nu}{2} (\det
\sigma+\det\sigma^\dagger)\nonumber\\
&&+\frac{\lambda_1}{2}\Bigg(tr(\sigma^\dagger\sigma)\Bigg)^2+\frac{\lambda_2}
{2}tr
\left(\sigma^\dagger\sigma-\frac{1}{2}tr(\sigma^\dagger\sigma)\right)^2+\dots
\end{eqnarray}
and we note that the axial $U(1)_A$-anomaly is reflected in the term
$\sim\nu$. The expectation value of $\sigma$ follows from the field
equation
\begin{equation}\label{A15}
\frac{\partial U}{\partial\sigma}=j^*=m_q.
\end{equation}
This suggests to define
\begin{equation}\label{A16}
U_j=U-tr(j^\dagger\sigma+\sigma^\dagger j)\qquad,\qquad
\frac{\partial U_j}{\partial\sigma}=0.
\end{equation}
In particular, for equal up-and-down-quark masses the minimum of $U_j$
occurs typically for
\begin{equation}\label{A17}
\langle\sigma\rangle=
\left(\begin{array}{lr}
\bar{\sigma}&\\
&\bar{\sigma}
\end{array}\right)
\end{equation}
and breaks the chiral $SU(2)_L\times SU(2)_R$ symmetry to a vectorlike
``diagonal'' $SU(2)_V$-symmetry. We recall that $\bar{\sigma}$ is
directly related to the quark-antiquark condensate
\begin{equation}\label{A18}
\bar{\sigma}=-\frac{1}{2}\langle\bar{\psi}\psi\rangle.
\end{equation}

One of the advantages of our formulation is the explicit chiral
symmetry of $U(\sigma)$ independently of the quark masses, since $m_q$
enters only through the source term in the field equation
(\ref{A15}). In particular, one can make direct contact to chiral
perturbation theory by representing the pseudoscalar (pseudo-)
Goldstone bosons by a nonlinear field\footnote{The coincidence of the
symbol $U$ for the nonlinear chiral field and  the effective potential
is unfortunate but kept here for the sake of agreement with widely
used conventions.} $U(x)$
\begin{equation}\label{A19}
\sigma(x)=\bar{\sigma}U(x)~,~U^\dagger U=1.
\end{equation}
Inserting the nonlinear field (\ref{A19}) into the effective potential
(\ref{A16}) yields
\begin{equation}\label{A20}
U_j(U)=\textup{const}-m_q\bar{\sigma} tr(U+U^\dagger)
-\frac{\nu}{2}\bar{\sigma}^2(\det U+\det U^\dagger).
\end{equation}
In the chiral limit $m_q=0$ the potential only involves the $\eta$
meson - the chiral $U(1)_A$-anomaly produces a mass term for the
$\eta$-meson. In this limit the pions are massless Goldstone
bosons. For nonvanishing quark masses also the pions acquire a mass
$m^2_\pi\sim m_q$.

Let us neglect the $\eta$-meson and discuss explicitly the
interactions of the pions
\begin{equation}\label{A21}
\det U=1~,~
U(x)=\exp\left\{\frac{i\vec{\tau}\vec{\pi}(x)}{f_\pi}\right\}.
\end{equation}
The kinetic term for $\sigma$ results in the nonlinear kinetic term
for the pions
\begin{equation}\label{A22}
\bar{Z}tr(\partial^\mu\sigma^\dagger\partial_\mu\sigma)
\rightarrow\bar{Z}\bar{\sigma}^2 tr (\partial^\mu U^\dagger \partial_\mu U)
=\frac{2\bar{Z}\bar{\sigma}^2}{f^2_\pi}\partial^\mu\vec{\pi}
\partial_\mu\vec{\pi}
+\dots
\end{equation}
where we identify the pion decay constant
\begin{equation}\label{A22a}
f_\pi=2\bar{Z}^{1/2}\bar{\sigma}.
\end{equation}
The effective action takes now the form familiar from chiral
perturbation theory
\begin{equation}\label{A24}
{\cal L}[U]=\frac{f^2_\pi}{4}\Big\{tr(\partial^\mu
U^\dagger\partial_\mu U) -2Bm_q tr (U+U^\dagger)+\dots\Big\}
\end{equation}
with
\begin{equation}\label{A25}
B=\frac{2\bar{\sigma}}{f^2_\pi}=-\frac{\langle\bar{\psi}\psi\rangle}{f^2_\pi}.
\end{equation}
In this treatment, however, ${\cal L}[U]$ is already the {\em
effective} Lagrangian from which the 1PI-vertices follow directly by
taking suitable derivatives with respect to the fields. No more
fluctuations have to be incorporated at this stage. In particular,
there are no explicit meson fluctuations.

In order to recover chiral perturbation theory we want to reformulate
our problem such that explicit pion fluctuations are incorporated in
the computation of the effective action. Indeed, it is possible to
reformulate the original functional integral for QCD (\ref{A1}) into
an exactly equivalent functional integral which now involves
explicitly an integration also over scalar fluctuations. This is
achieved by means of a Hubbard-Stratonovich transformation
\cite{HS}. Let us denote the scalar fermion bilinears by
\begin{equation}\label{A26}
\tilde{\sigma}_{ab}=\bar{\psi}_{Lb}\psi_{Ra}~,~
\langle\tilde{\sigma}_{ab}\rangle=\sigma_{ab}.
\end{equation}
We next introduce a unit into the functional integral (\ref{A1})
\begin{equation}\label{A27}
{\mathbbm 1}= N\int{\cal D}\sigma'\exp
\left\{-\int\frac{d^4q}{(2\pi)^4}tr\Big\{(\sigma'-\lambda_\sigma
\tilde{\sigma}-j)^\dagger
\lambda_\sigma^{-1}(\sigma'-\lambda_\sigma\tilde{\sigma}-j)\Big\}\right\}
\end{equation}
where $N$ is an irrelevant normalization constant. The functional
integral for the partition function $Z[j]$ involves now an additional
integration over the scalar field $\sigma'$
\begin{equation}\label{XXA}
Z[j]=\int{\cal D}\psi{\cal D}A{\cal D}\sigma' ~\exp(-S^{(\sigma)}_j).
\end{equation}
Here the action $S_j$ in eq. (\ref{A4}) is replaced by
$S^{(\sigma)}_j$, i.e.
\begin{eqnarray}\label{A28}
S^{(\sigma)}_j&=&S_0+\int\frac{d^4q}{(2\pi)^4}tr\Big\{\sigma^{\prime\dagger}
\lambda^{-1}_\sigma\sigma'
-(\sigma^{\prime\dagger}\tilde{\sigma}+\tilde{\sigma}^\dagger\sigma')+\lambda_\sigma
\tilde{\sigma}^\dagger\tilde{\sigma}\nonumber\\
&&-(j^\dagger\lambda^{-1}_\sigma\sigma'+\sigma^{\prime\dagger}\lambda^{-1}_\sigma
j)+ j^\dagger\lambda^{-1}_\sigma j\Big\}
\end{eqnarray}
where $S_0$ is the QCD-action for quarks and gluons without the quark
mass term. Indeed, the quark mass term $\sim
tr(j^\dagger\widetilde{\sigma})$ is canceled by a corresponding term
from eq. (\ref{A27}) such that the nonvanishing quark masses appear
now as a source term multiplying the scalar field $\sigma'$.

In eq. (\ref{A27}) $\sigma'$ stands for $\sigma'_{ab}(q)$ and
$\lambda^{-1}_\sigma(q)$ is an arbitrary positive function of $q^2$
such that the Gaussian integral is well defined and rotation symmetry
preserved. If we choose
\begin{equation}\label{A29}
\lambda_\sigma=\frac{1}{M^2+Zq^2}
\end{equation}
we can identify the term
$\sigma^{\prime\dagger}\lambda_\sigma^{-1}\sigma'$ with a kinetic and
mass term for $\sigma'$ such that $\lambda_\sigma$ corresponds to the
``classical scalar propagator''. The term
$\tilde{\sigma}^\dagger\sigma'$ denotes a Yukawa coupling of the
$\sigma'$ field to the quarks. Finally, the expression
$\lambda_\sigma\tilde{\sigma}^\dagger\tilde{\sigma}$ denotes a four
quark interaction that is nonlocal for $Z\neq 0$.

The explicit four-quark-interaction is cumbersome and we may wish to
omit it. This can be done if we introduce into the original action for
quarks and gluons an additional four-quark-interaction with the
opposite sign - this is then canceled by the piece arising from the
Hubbard-Stratonovich transformation. At first sight, this seems to be
a high prize to pay since  we are not dealing any more with the
standard QCD action where the quarks interact only via gluon
exchange. With a second look, however, this is no
problem. Implementing this modified QCD action for lattice simulations
 will even result in an ``improved'' QCD action if suitable values for
  $M^2$ and $Z$ are chosen in eq. (\ref{A29}).

The basic ingredient for this argument is universality. In QCD the
precise form of the short distance action (classical action) is
actually irrelevant, provided this action belongs to the universality
class of QCD. Independently of the precise form of the classical
action the long distance behavior will then only depend on the value
of the renormalized gauge coupling (which sets the scale via
dimensional transmutation) and the quark masses. This fact is widely
used in the form of ``improved actions'' for lattice QCD. In a
qualitative sense the mapping of the continuous QCD onto a lattice
with finite lattice spacing $a$ involves an integration of
fluctuations with momenta $q^2>\pi^2/a^2$. Performing the gluon
integration in perturbation theory the box diagrams with exchange of
two gluons indeed lead to four-quark-interactions in the form needed
here  (plus others) \cite{Meg,Gi}. Their strength can be used to
optimize $M^2$ and $Z$ for a given lattice spacing $a$, thus improving
the action. We emphasize, however,  that due to universality this
improvement is not necessary, in principle. The precise values of
$M^2$ and $Z$ should not be important as long as $M^2$ is not too
small and $Z$ not too large  such that the interaction remains
``essentially local'' and within the universality class of QCD.

In the formulation with additional scalars the issue of the correct
universality class actually needs some consideration. For very small
$M^2$ or very large quartic coupling $\lambda_\sigma$ (in units of
$a$) the theory will be in a different universality class where the
scale of chiral symmetry breaking is set by $\lambda_\sigma$ and not
by $\Lambda_{QCD}$  as appropriate for QCD. This issue is discussed
quantitatively in \cite{Gi}. In particular, for too small $M^2$ the
model contains an additional relevant parameter beyond the gauge
coupling (and quark masses).\footnote{This also tells us that QCD
  lattice simulations with explicit scalar fields are only meaningful
  with dynamical quarks. In the quenched approximation the scalar
  sector decouples from the gluons and therefore induces the relevant
  parameters of a pure scalar theory that should not be present for
  the QCD universality class.}

We suggest that lattice simulations with explicit meson degrees of
freedom and the action (\ref{A28}) (without the quartic term $\sim
\lambda_\sigma \tilde\sigma^\dagger\tilde\sigma$ and for appropriate
$M^2,Z$) may actually permit a direct access to the chiral limit of
vanishing quark masses. In the current formulation (without explicit
meson degrees of freedom) one of the main obstacles for the simulation
of the chiral limit is the complicated spectrum of the Dirac operator
for the quarks. In presence of meson fields the dominant
configurations would correspond to non-zero values of $\sigma'$. In
term, this induces a mass gap for the Dirac operator due to the Yukawa
coupling  $\tilde\sigma^\dagger\sigma'$. This should make simulations
with dynamical quarks much simpler. In addition, the meson
fluctuations provide the degrees of freedom appearing in chiral
perturbation in a very explicit way.  The contact to chiral
perturbation theory should therefore be facilitated.

Our formulation where the source term appears  explicitly in the transformation
(\ref{A27})  has the additional advantage that the quark
masses only appear as terms linear in the scalar field. On the  level
of the effective action we can perform a single computation for all
values of the quark masses and fully  exploit the chiral symmetry. The
quark masses appear then only as source terms in the field
equations. Still, we  have to take into account the modifications due
to the term $j^\dagger\lambda^{-1}_\sigma j$ in eq. (\ref{A28}). This
modifies the relation between the expectation value of $\sigma'$ and
$\tilde\sigma$ according to 
\begin{equation}\label{A30}
\frac{\delta W}{\delta j^*}=\langle\lambda^{-1}_\sigma
(\sigma'-j)\rangle=\langle\tilde{\sigma}\rangle=\sigma
\end{equation}
such that
\begin{equation}\label{A31}
\varphi=\langle\sigma'\rangle=\lambda_\sigma\sigma+j.
\end{equation}
We may bring the functional  integral even closer to the standard
formulation for a theory with quarks, gluons and scalars by using
standard sources of $\sigma'$
\begin{equation}\label{33A}
j_\varphi=\lambda^{-1}_\sigma j
\end{equation}
and subtracting the term quadratic in $j$ which does not involve the fields
\begin{equation}\label{A32}
S^{(\varphi)}_j=S^{(\sigma)}_j-
\int j^\dagger\lambda^{-1}_\sigma j.
\end{equation}
In this formulation, the generating functional for the connected
Greens functions is related to eq. (\ref{A7})  by
\begin{equation}\label{A33}
W_\varphi=W+\int j^\dagger\lambda^{-1}_\sigma j
\end{equation}
and the effective action is defined as usual
\begin{equation}\label{A34}
\Gamma[\varphi]=-W_\varphi+\int(j^\dagger_\varphi\varphi+\varphi^\dagger 
j_\varphi).
\end{equation}
Comparing with $\Gamma[\sigma]$ (\ref{A10}) one obtains
\begin{equation}\label{A35}
\Gamma[\varphi]=\Gamma[\sigma]+\int\frac{\delta\Gamma}{\delta\sigma}
\lambda^{-1}_\sigma
\frac{\delta\Gamma}{\delta\sigma}.
\end{equation}
(We have not used different symbols, but it should be clear that
$\Gamma[\varphi]$ and $\Gamma[\sigma]$ are not related simply by a
variable transformation.)
The difference between $\Gamma[\varphi]$ and $\Gamma[\sigma]$ vanishes
at the extrema of $\Gamma[\sigma]$ but modifies, for example, the
behavior for large values of the fields. Typically, $\Gamma[\varphi]$
is a better behaved quantity. For example, the effective potential is
bounded from below which is not guaranteed for $\Gamma[\sigma]$. The 
field equations derived
from $\Gamma[\varphi]$ take the standard form
\begin{equation}\label{A36}
\frac{\delta\Gamma}{\delta\varphi}=j^\dagger_\varphi=\lambda^{-1}_\sigma 
j^\dagger.
\end{equation}

\section{QCD effective action with composite scalar fields}
\label{QCDeffectiveaction}
Let us next discuss the effective action for QCD in a formulation
where explicit scalar fields  for composite $\bar{q}q$ bilinears are
used according to the preceding section. We will see that a very
simple form of the effective action can account for a great deal of
the strong interaction phenomenology in a surprisingly simple fashion.
By de\-finition the effective action generates the
one-particle-irreducible (1PI) correlation functions and includes all
quantum fluctuations. It therefore contains the direct information
about the propagators and proper vertices.

In addition to the quark and gluon fields we consider scalar fields
with the transformation properties of quark-antiquark pairs. With
respect to the color and chiral flavor rotations $SU(3)_C\times
SU(3)_L\times SU(3)_R$ the three light left-handed and right-handed
quarks $\psi_L,\psi_R$ transform as (3,3,1) and (3,1,3),
respectively. Quark-antiquark bilinears therefore contain a color
singlet $\Phi(1,\bar 3, 3)$ and a color octet $\chi(8,\bar 3, 3)$
\begin{eqnarray}\label{2.1}
\gamma_{ij}&=&\chi_{ij}+\frac{1}{\sqrt3}\phi\ \delta_{ij},\nonumber\\
\phi&=&\frac{1}{\sqrt3}\gamma_{ii}\quad,\quad \chi_{ii}=0.
\end{eqnarray}
Here we use a matrix notation for the flavor indices and write the
color indices $ i,j$ explicitly, e.g. $\chi _{ij,ab}\equiv\chi_{ij},\
\phi_{ab}
\equiv\phi$. Then $\gamma_{ij}$ contains 81 complex scalar fields.
Similarly, the quark fields are represented as three flavor vectors
$\psi_{ai}\equiv\psi_i$, $\bar\psi_{ia}\equiv\bar\psi_i$. As usual we
represent the eight $SU(3)_C$-gauge fields by
\begin{equation}\label{2.5}
A_{ij,\mu}=\frac{1}{2}A^z_\mu(\lambda_z)_{ij}
\end{equation}
where $\lambda_z$ are the eight Gell-Mann matrices normalized
according to ${\rm Tr}(\lambda_y\lambda_z)=2\delta_{yz}$.

We consider a very simple  effective Lagrangian containing only terms
with dimension up to  four\footnote{For our Euclidean conventions for fermions
see
\cite{Conv}.}
\begin{eqnarray}\label{2.6}
{\cal L}&=&iZ_\psi\bar\psi_i\gamma^\mu\partial_\mu\psi_i+g
Z_\psi\bar\psi_i
\gamma^\mu A_{ij,\mu}\psi_j+\frac{1}{2}G^{\mu\nu}_{ij}
G_{ji,\mu\nu}\nonumber\\ &&+{\rm Tr}\{(D^\mu\gamma_{ij})^\dagger D_\mu
\gamma_{ij}\}+U(\gamma)
\nonumber\\
&&+Z_\psi\bar\psi_i[(h\phi\delta_{ij}+\tilde h\chi_{ij})
\frac{1+\gamma_5}{2}-(h\phi^\dagger\delta_{ij}+\tilde h
\chi^\dagger_{ji})\frac{1-\gamma_5}{2}]\psi_j.
\end{eqnarray}
Here $G_{ij,\mu\nu}=\partial_\mu A_{ij,\nu}-\partial_\nu A_{ij,\mu}-ig
A_{ik,\mu}A_{kj,\nu}+ig A_{ik,\nu}A_{kj,\mu}$ and the interaction
between gluons and $\chi$ arise from the covariant derivative
\begin{equation}\label{2.6a}
D_\mu\gamma_{ij}=\partial_\mu\gamma_{ij}- ig
A_{ik,\mu}\gamma_{kj}+ig\gamma_{ik}A_{kj,\mu}.
\end{equation}
In our notation the transposition acts only on flavor indices,
e.g. $(\gamma^\dagger _{ij})_{ab}=\gamma^*_{ij,ba}$.  The effective
potential
\begin{eqnarray}\label{2.7}
&&U(\gamma)=U_0(\chi,\phi)-\frac{1}{2}\nu (\det\ \phi+\det
\phi^\dagger) -\frac{1}{2}\nu'
(E(\phi,\chi)+E^*(\phi,\chi))\nonumber\\
&&E(\phi,\chi)=\frac{1}{6}\epsilon_{a_1a_2a_3}\epsilon_{b_1b_2b_3}
\phi_{a_1b_1}\chi_{ij,a_2b_2}\chi_{ji,a_3b_3}
\end{eqnarray}
conserves the symmetry $SU(3)_c\times SU(3)_L\times SU(3)_R$. It is
decomposed into  a part $U_0$ which conserves also the axial $U(1)_A$
symmetry and an anomaly contribution. The  latter is parameterized
here by the 't Hooft terms \cite{Anom} $\sim\nu,\nu'$. The details of
$U$ will not be relevant for our discussion except for the location of
the minimum that will be characterized by two parameters $\sigma_0$
and $\chi_0$.  Finally, explicit chiral symmetry breaking is induced
by a linear term \cite{6}
\begin{eqnarray}\label{2.8}
{\cal L}_j&=&-\frac{1}{2}Z^{-1/2}_\phi\ {\rm Tr}\
(j^\dagger\phi+\phi^\dagger   j)\nonumber\\ j=j^\dagger&=&a_q\bar
m=a_q\  diag(\bar m_u,\bar m_d,\bar m_s)
\end{eqnarray}
with $\bar m_q$ the current quark masses normalized at some
appropriate scale, say $\mu=2$ GeV. We note that the quark wave
function renormalization $Z_\psi$ can be absorbed by a rescaling of
$\psi$. We keep it here and normalize $\psi$ at a perturbative scale
$\mu=2$ GeV.

In general, the wave function renormalizations $Z_\phi,Z_\psi$ and the
dimensionless  couplings $g,h$ depend on momentum or an appropriate
renormalization scale $\mu$. The  phenomenological analysis will
mainly concentrate on the low energy limit where we take all
couplings as constants.  Besides the explicit chiral symmetry breaking
in (\ref{2.8}) the phenomenology will be described in terms of the
seven real parameters $g,h,\tilde h,\sigma_0,\chi_0, \nu$  and
$\nu'$. Here $\nu$ and $\nu'$ will only enter the mass of $\eta'$ and
$h,\tilde{h}$ will be fixed  by the baryon masses. Then three
parameters $g,\sigma_0,\chi_0$ remain for the description of the
pseudoscalars and vector mesons and their interactions, including the
interactions with baryons.  In fact, the effective action (\ref{2.6})
is the most general\footnote{The only exception is the omission of a
possible $U(1)_A$-violating term cubic in $\chi$ which can be found in
\cite{JW}.}  one which contains only ``renormalizable''
interactions and is consistent with $SU(3)_C\times SU(3)_L\times
SU(3)_R$ symmetry, as well as with the discrete transformations parity
and charge conjugation.

We will show that the masses and interactions of the octet
of baryons as well as the pseudoscalar and vector mesons are indeed well
described by the simple effective action (\ref{2.6}), (\ref{2.8}). We
find this rather remarkable in view of the fact that we have only
added rather straightforward terms for composite scalars (``scalar
Vervollst\"andigung'' \cite{CWSSB}). Of course, the crucial ingredient
is the nontrivial minimum of $U(\chi,\phi)$.

Actually, there is no reason why only interactions involving  scalar and
pseudoscalar quark-antiquark bilinears should be  present as in
(\ref{2.6}). Effective four-quark interactions (1PI) in the vector or
axial-vector channel are certainly induced by fluctuations, and we
will discuss them below.  There is also no need that the
effective action has to be of the ``renormalizable form''
(\ref{2.6}). In fact, we have at present no strong argument why
higher-order operators have to be small.  In the spirit that a useful
dual description should not be too complicated, we simply investigate
in this paper to what extent the effective action
(\ref{2.6}) is compatible with observation.  If
successful, the neglected subleading terms may be considered later for
increased quantitative accuracy. Furthermore, we know that QCD
contains higher resonances like the $\Delta$ or the axial-vector
mesons. They are not described by  the  effective action
(\ref{2.6}). One may include them by the introduction of additional
fields for bound states.  This is, however, not the
purpose of the present paper. We only mention here that ``integrating
out'' the missing resonances with lowest mass presumably gives the
leading contribution to the neglected higher-order operators. Those
will typically contain ``nonlocal behavior'' in the momentum range
characteristic for the resonances.

For high enough momenta (say $q^2>(2GeV)^2$) our description should
coincide with  perturbative QCD. Obviously, in this momentum range
$g(q)$ should coincide with the perturbative gauge coupling, providing
for a ``matching'' of couplings.  Furthermore, the (tree) exchange of
the scalars induces an effective four-quark-interaction. In the
perturbative QCD picture this matches with (part of) the
1PI-four-quark-interaction generated by ``box diagrams'' with gluon
exchange.

The essential point of our picture of the QCD-vacuum concerns the
location of the minimum of the effective potential. We will
concentrate here on two directions in field space according to
\begin{equation}\label{2.17a}
\phi_{ab}=\bar{\sigma}\delta_{ab},
\end{equation}
\begin{equation}\label{2.17}
\chi_{ij,ab}=\frac{1}{\sqrt{24}}\bar{\chi}\lambda^z_{ji}\lambda^z_{ab}
=\frac{1}{\sqrt6}\bar{\chi}(\delta_{ia}\delta_{jb}-\frac{1}{3}
\delta_{ij}\delta_{ab})
\end{equation}
with real fields $\bar{\sigma},\bar{\chi}$. Our central assumption
states that the minimum of $U(\bar{\sigma},\bar{\chi})$ occurs for
nonvanishing
\begin{equation}\label{2.17b}
\bar{\sigma}=\sigma_0~,~\bar{\chi}^2=\chi^2_0.
\end{equation}
In this context we emphasize that the source term (\ref{2.8})
preserves the $SU(3)_c$ color symmetry. The minimum at
$\bar{\chi}^2=\chi^2_0$ therefore corresponds to a whole orbit in the
field space of $\chi_{ij,ab}$. This orbit obtains by applying
$SU(3)_c$-transformations on the configuration (\ref{2.17}). In
consequence, our statement about the minimum of the effective
potential is perfectly gauge invariant! In contrast, for nonzero quark
masses the minimum in the direction of $\phi_{ab}$,
i.e. $\phi_{ab}=\sigma_0\delta_{ab}$, is unique. As it stands, there
exists a simple argument in QCD that the effective potential for color
non singlet fields has its minimum at zero, i.e. $\chi_0=0$. Our
investigation for $\chi_0\not=0$ can therefore only hold in a gauge
fixed version or after a suitable redefinition of $\chi$ by a
multiplication with phase factors that makes it formally gauge invariant.

\section{Higgs picture}
\label{higgspicture}
For most of the discussion of this paper we will employ a gauge
invariant description which only uses the assumption that the minimum
of $U$ occurs for nonzero $\bar{\chi}$, without fixing a particular
direction in the orbit of minima. Nevertheless, the most striking
features of our picture of the QCD-vacuum become easily apparent if we
fix for a moment the direction in field space. In this section we
describe this ``Higgs picture'' where the vacuum is characterized by a
fixed field
\begin{equation}\label{2.17c}
\langle\chi_{ij,ab}\rangle=\frac{1}{\sqrt{24}}\chi_0\lambda^z_{ji}
\lambda^z_{ab}
=\frac{1}{\sqrt6}\chi_0(\delta_{ia}\delta_{jb}-\frac{1}{3}
\delta_{ij}\delta_{ab}).
\end{equation}
This expectation value is invariant under a combination of $SU(3)_c$
transformations and vectorlike flavor transformations. More precisely,
the symmetry $SU(3)_V$ of the vacuum consists of those transformations
where identical left and right flavor rotations and {\em transposed}
color rotations are performed with opposite angles.  With respect to
the {\em transposed} color rotations the quarks behave as
antitriplets.  In consequence, under the combined transformation they
transform as an octet plus a singlet, with all quantum numbers
identical to the baryon octet/singlet.  In particular, the electric
charges, as given by the generator $Q=
\frac{1}{2}\lambda_3+\frac{1}{2\sqrt3}\lambda_8$ of $SU(3)_V$, are integer.
We therefore identify the quark field with the lowest baryon octet and
singlet -- this is quark-baryon duality. Similarly, the gluons
transform as an octet of vector-mesons, again with the standard
charges for the $\rho, K^*$ and $\omega$ mesons.  We therefore
describe these vector mesons by the gluon field -- this is gluon-meson
duality.

Due to the Higgs mechanism all gluons acquire a mass. For equal quark
masses conserved $SU(3)_V$ symmetry implies that all masses are
equal. They should be identified with the average mass of the vector
meson multiplet $\bar M_\rho^2=\frac{2}{3}
M^2_{K^*}+\frac{1}{3}M^2_\rho=(850\ {\rm MeV})^2$. One finds
\begin{equation}\label{2.18a}
\bar M_\rho=g|\chi_0|=850\ {\rm MeV}.
\end{equation}
The breaking of chiral symmetry by $\sigma_0$ and $\chi_0$ also
induces masses for the baryon octet and singlet.  The singlet mass is
larger than the octet mass (see sect. 5).

In order to get familiar with the Higgs picture of QCD, it seems
useful to understand in more detail the consequences of spontaneous
color symmetry breaking for the electromagnetic interactions of
hadrons. We start by adding to the effective Lagrangian (\ref{2.6}) a
coupling to a $U(1)$-gauge field $\tilde B_\mu$ by making derivatives
covariant
\begin{eqnarray}\label{3.1}
D_\mu\psi&=&(\partial_\mu-i\tilde e\tilde Q\tilde
B_\mu)\psi\nonumber\\
D_\mu\gamma_{ij}&=&\partial_\mu\gamma_{ij}-i\tilde e[
\tilde Q,\gamma_{ij}]\tilde B_\mu-...
\end{eqnarray}
with $\tilde Q=\frac{1}{2}\lambda_3+\frac{1}{2\sqrt3}\lambda_8
=diag(\frac{2}{3},-\frac{1}{3}, -\frac{1}{3})$ acting on the flavor
indices. Furthermore we supplement the Maxwell term
\begin{equation}\label{3.2}
{\cal L}_{\tilde B}=\frac{1}{4}\tilde B^{\mu\nu}\tilde B_{\mu\nu}
\ , \ \tilde B_{\mu\nu}=\partial_\mu\tilde B_\nu-\partial_\nu
\tilde B_\mu.
\end{equation}
Whereas the quarks carry fractional $\tilde Q$, the abelian charges of
the scalars are integer. In particular, the expectation value $<\phi>$
is neutral, $[\tilde Q,<\phi>]=0$, whereas some components of
$<\chi_{ij,ab}>$ carry charge, namely for $k=2,3$
\begin{eqnarray}\label{3.3}
{}[\tilde Q,<\chi_{1k,1k}>]&=&<\chi_{1k,1k}>\nonumber\\ {}[\tilde
Q,<\chi_{k1,k1}>]&=&-<\chi_{k1,k1}>.
\end{eqnarray}
The expectation value (\ref{2.17}) therefore also breaks the local
$U(1)$ symmetry associated with $\tilde Q$. The abelian color charge
$(Q_C)_{ij}=\frac{1}{2}
(\lambda_3)_{ij}+\frac{1}{2\sqrt3}(\lambda_8)_{ij}$ of these fields
is, however, equal to $\tilde Q$. In consequence, a local abelian
symmetry with generator $\tilde Q-Q_C$ remains unbroken
\begin{equation}\label{3.4}
\tilde Q_{ac}<\chi_{ij,cb}>-<\chi_{ij,ac}>\tilde Q_{cb}-
(Q_C)_{il}<\chi_{lj,ab}>+<\chi_{il,ab}>(Q_C)_{lj}=0.
\end{equation}
The corresponding gauge field corresponds to the photon.

The situation encountered here is completely analogous to the Higgs
mechanism in electroweak symmetry breaking. The mixing between the
hypercharge boson and the $W_3$ boson in the electroweak theory
appears here as a mixing between $\tilde B_\mu$ and a particular gluon
field $\tilde G_\mu$ which corresponds to $A_{ij,\mu}=\frac{\sqrt
3}{2}(Q_C)_{ij}\tilde G_\mu$. Let us restrict the discussion to the
gauge bosons $\tilde B_\mu$ and $\tilde G_\mu$. Then the covariant
derivative for fields with a fixed value of $\tilde Q$ and $Q_C$ is
given by
\begin{equation}\label{3.5}
D_\mu=\partial_\mu-i\tilde e\tilde B_\mu \tilde Q-i\tilde g\tilde
G_\mu Q_C
\end{equation}
with $\tilde g=\frac{\sqrt3}{2}g$. Due to the ``charged'' expectation
values (\ref{3.3}) a linear combination of $\tilde B_\mu$ and $\tilde
G_\mu$ gets massive, as can be seen from the quadratic Lagrangian
\begin{equation}\label{3.6}
{\cal L}^{(2)}_{em}=\frac{1}{4}\tilde B^{\mu\nu}\tilde B_{\mu\nu}
+\frac{1}{4}\tilde G^{\mu\nu}\tilde G_{\mu\nu}+
\frac{2}{3}\chi^2_0(\tilde g\tilde G^\mu+\tilde e\tilde B^\mu)
(\tilde g\tilde G_\mu+\tilde e\tilde B_\mu).
\end{equation}
The massive neutral vector meson $R_\mu$ and the massless photon
$B_\mu$ are related to $\tilde G_\mu$ and $\tilde B_\mu$ by a mixing
angle
\begin{eqnarray}\label{3.7}
R_\mu&=&\cos\theta_{em}\tilde G_\mu+\sin\theta_{em}\tilde
B_\mu\nonumber\\ B_\mu&=&\cos\theta_{em}\tilde
B_\mu-\sin\theta_{em}\tilde G_\mu\nonumber\\
tg\theta_{em}&=&\frac{\tilde e}{\tilde g}
\end{eqnarray}
and we note that the mass of the neutral vector meson is somewhat
enhanced by the mixing
\begin{equation}\label{3.8}
M_{V_0}=g\chi_0/\cos \theta_{em}.
\end{equation}
The mixing is, however, tiny for the large value $\bar\alpha
_s=g^2/4\pi\approx 3$ that we will find below. In terms of the mass
eigenstates the covariant derivative (\ref{3.5}) reads now
\begin{equation}\label{3.9}
D_\mu=\partial_\mu-ieQB_\mu-i\tilde g\cos\theta_{em}(Q_C
+tg^2\theta_{em}\tilde Q)R_\mu
\end{equation}
and we observe the universal electromagnetic coupling
\begin{equation}\label{3.10}
e=\tilde e\cos\theta_{em}
\end{equation}
for all particles with electric charge $Q=\tilde Q-Q_C$.  This coupling
is exactly the same for the colored quarks and the colorless leptons
as it should be for the neutrality of atoms.  For an illustration we
show the charges $Q_C,\tilde Q$ and $Q$ for the nine light quarks in
table 1.

\begin{center}
\begin{tabular}{l|ccc|c|}
&$\tilde Q$&$Q_c$&$Q$&\\
\hline
$u_1$&$2/3$&$2/3$&0&$\Sigma^0,\Lambda^0,S^0$\\
$u_2$&$2/3$&$-1/3$&1&$\Sigma^+$\\ $u_3$&$2/3$&$-1/3$&1&$p$\\
\hline
$d_1$&$-1/3$&$2/3$&-1&$\Sigma^-$\\
$d_2$&$-1/3$&$-1/3$&0&$\Sigma^0,\Lambda^0,S^0$\\
$d_3$&$-1/3$&$-1/3$&0&$n$\\
\hline
$s_1$&$-1/3$&$2/3$&-1&$\Xi^-$\\ $s_2$&$-1/3$&$-1/3$&0&$\Xi^0$\\
$s_3$&$-1/3$&$-1/3$&0&$\Lambda^0,S^0$
\end{tabular}\\
\end{center}

\medskip
{\em Table 1}: Abelian charges of quarks and association with
baryons. The baryon  singlet is denoted by $S^0$.

\bigskip

The Higgs picture (with fixed direction of $<\chi>$) yields a
particularly simple description of the mass generation for the gluons
and the charges of the physical excitations. It has, however, some
shortcomings concerning the simplicity of the description of the meson
interactions and the correct interpretation of the global baryon
number of the excitations \cite{CWSSB}. These issues can be dealt with
in a much easier way in the gauge invariant description.

\section{Nonlinear fields}
\label{nonlinearfields}
Let us now turn back to a gauge invariant description and use
appropriate nonlinear fields. We will recover all the findings of the
``Higgs picture'' in the preceding section, now in a gauge invariant
way. Indeed, we never need an assumption that an octet condensate
occurs in a particular direction in field space (which would
correspond to ``spontaneous symmetry breaking'' of the color
symmetry). Furthermore, the interactions of the light mesons are most
easily described in the gauge-invariant picture, using nonlinear
fields\footnote{On the level of the effective action the quantum
fluctuations are already included. Therefore Jacobians for nonlinear
field transformations play no role. All ``coordinate choices'' in the
space of fields are equivalent.}. For the ``Goldstone directions'' we
introduce unitary matrices $W_L, W_R, v$ and define
\begin{eqnarray}\label{4.1}
\psi_L&=&Z_\psi^{-1/2}W_LN_Lv\ ,\ \psi_R=Z_\psi^{-1/2}W_RN_Rv,\nonumber\\
\bar\psi_L&=&Z_\psi^{-1/2}v^\dagger\bar N_LW^\dagger_L\ , \ \bar\psi_R
=Z_\psi^{-1/2}v^\dagger\bar N_RW_R^\dagger\nonumber\\
A_\mu&=&-v^TV^T_\mu v^*-\frac{i}{g}\partial_\mu v^Tv^*\ ,\
U=W_RW_L^\dagger,\nonumber\\
\phi&=&W_RSW^\dagger_L\ , \ \chi_{ij,ab}=(W_R)_{ac}v^T_{ik}
X_{kl,cd}v^*_{lj}(W^\dagger_L)_{db}.
\end{eqnarray}
Here we have again extended the matrix notation to the color indices
with the quark/baryon nonet represented by a complex $3\times 3$
matrix $\psi\equiv\psi_{ai},\bar\psi\equiv\bar\psi_{ia}$. The
decomposition of $\phi$ is such that $S$ is a hermitean matrix. The
restrictions on $X$, which are necessary to avoid double counting,
play no role in the present work.  Besides $v$ all nonlinear fields
are color singlets. The field $v^\dagger$ transforms as a color
anti-triplet similar to a quark-quark pair or diquark.  By virtue of
triality - the center elements of global $SU(3)_c$ and $SU(3)_L\times
SU(3)_R$ are identified with particular global transformations of the
$U(1)_B$-symmetry which corresponds to conserved baryon number - the
fields belonging to color representations in the class of the
antitriplets carry baryon number $2/3~(mod~1)$. We therefore assign to
$v^\dagger$ (or the product $W^\dagger v^\dagger)$
a nonvanishing baryon number $B=2/3$ (as appropriate for a diquark). In
consequence, the baryons $N\sim W^\dagger\psi v^\dagger$ transform as color
singlets with baryon number $B=1$, as it should be.

The nonlinear fields (\ref{4.1}) are adapted to the physical
excitations around the minimum of the effective potential. We
encounter the gauge singlet baryon field $N$ and the vector-meson
fields $V_\mu$. The pseudoscalar mesons correspond to the (pseudo-)
Goldstone bosons described by $U=W_RW_L^\dagger$. We will see below
that the remaining nonlinear degrees of freedom contained in $v$ and
$W_{L,R}$ are  gauge degrees of freedom. The effective action
(\ref{2.6}) can now be expressed in terms of the physical fields
$N,V^\mu,U$ by inserting the definitions (\ref{4.1}). In this form it
will be straightforward to extract the physical excitations and their
interactions.

Before proceeding with this program, we may verify that the effective
action is invariant under the following field transformations
\begin{eqnarray}\label{4.2}
\delta W_L&=&i\Theta_LW_L-iW_L\Theta_P\ ,\ \delta W_R=i\Theta_R
W_R-iW_R\Theta_P\ ,\nonumber\\
\delta v&=&i\Theta_P v+iv\Theta_C^T\ ,\ \delta U=i\Theta_R
U-iU\Theta_L
\nonumber\\
\delta N_L&=&i[\Theta_P,N_L]\ ,\ \delta N_R=i[\Theta_P,N_R]\ ,\
\delta V_\mu=i[\Theta_P,V_\mu]+\frac{1}{g}\partial_\mu\Theta_P,
\nonumber\\
\delta S&=&i[\Theta_P,S]\ ,\ (\delta X)_{ij,ab}=i(\Theta_P)_{ac}X_{ij,cb}-iX
_{ij,ac}(\Theta_P)_{cb}\nonumber\\ &&\qquad\qquad\qquad\qquad\ \
-i(\Theta_P^T)_{ik}X_{kj,ab}+iX_{ik,ab} (\Theta_P^T)_{kj}.
\end{eqnarray}
Here $\Theta_c,\Theta_L,\Theta_R$ are special unitary $3\times 3$
matrices corresponding to $SU(3)_c,SU(3)_L$ and $SU(3)_R$,
respectively. (We have no displayed additional global symmetries like
the one corresponding to baryon number.) As a consequence of the
nonlinear formulation one observes the appearance of a new local
symmetry $U(3)_P= SU(3)_P\times U(1)_P$ under  which the nucleons $N$
and the vector meson fields  $V_\mu$ transform as octets and singlets
\begin{equation}\label{4.2AA}
\Theta_P=\frac{1}{2}(\theta_P^z(x)\lambda_z+\theta^0_P(x)\lambda_0)\ ,\
\lambda_0\equiv\frac{2}{\sqrt6}.
\end{equation}
This symmetry acts only on the nonlinear fields whereas $\psi, A_\mu,
\phi$ and $\chi$ are invariant. It reflects the possibility of local
reparameterizations of the nonlinear fields. It will play the role of
the hidden gauge symmetry underlying vector dominance.

For the present investigation we omit the scalar excitations except
for the ``Goldstone directions'' contained in $W_L, W_R, v$. We can
therefore replace $X_{kl,ab}$ by the expectation value (\ref{2.17})
and use $<S>=\sigma_0$ such that
\begin{equation}\label{4.3}
\Phi=\sigma_0U\quad,\quad
\chi_{ij,ab}=\frac{1}{\sqrt6}\chi_0\{(W_Rv)_{ai}(v^\dagger
W^\dagger_L)_{jb} -\frac{1}{3}U_{ab}\delta_{ij}\}.
\end{equation}
In terms of the nonlinear field coordinates the Lagrangian (\ref{2.6})
reads
\begin{eqnarray}\label{4.4}
{\cal L}&=&Tr\{i\bar N\gamma^\mu(\partial_\mu +iv_\mu-i\gamma^5
a_\mu)N-g\bar N\gamma^\mu NV_\mu
\}\nonumber\\
&&+\frac{1}{2}Tr\{V^{\mu\nu} V_{\mu\nu}\} +g^2\chi^2_0\ Tr\ \{\tilde
V^\mu \tilde V_\mu\}
+(\frac{2}{9}\nu'\chi_0^2\sigma_0-\nu\sigma^3_0)\cos\theta\nonumber\\
&&+( h\sigma_0-\frac{\tilde h}{3\sqrt6}\chi_0)
\ Tr\{\bar N \gamma_5N\}+\frac{\tilde h}{\sqrt6}\chi_0\ Tr\ \bar
N\gamma^5\ 
Tr\  
N\nonumber\\ &&+(\sigma^2_0+\frac{7}{36}\chi_0^2)\ Tr\ \{\partial^\mu
U^\dagger
\partial_\mu  
U\}+\frac{1}{12}\chi^2_0
\ \partial^\mu\theta\ \partial_\mu\theta\nonumber\\
&&+\chi_0^2\ Tr\{\tilde v^\mu\tilde v_\mu\}+2g\chi^2_0\ Tr\{\tilde V
^\mu\tilde v_\mu\}\end{eqnarray} with
\begin{eqnarray}\label{4.4a}
&&v_\mu=-\frac{i}{2}(W^\dagger_L\partial_\mu W_L+W_R^\dagger
\partial_\mu W_R),\nonumber\\
&&a_\mu=\frac{i}{2}(W_L^\dagger\partial_\mu
W_L-W_R^\dagger\partial_\mu W_R) =-\frac{i}{2}W_R^\dagger \partial_\mu
UW_L,\nonumber\\ &&Tr\ v_\mu=-\frac{i}{2}\partial_\mu(\ln \det\
W_L+\ln\det W_R),
\nonumber\\
&&Tr\ a_\mu=-\frac{i}{2}Tr\ U^\dagger\partial_\mu
U=-\frac{1}{2}\partial_\mu\theta,\nonumber\\ &&V_{\mu\nu}=\partial_\mu
V_\nu-\partial_\nu V_\mu \ -ig[V_\mu,V _\nu]=\partial_\mu\tilde
V_\nu-\partial_\nu\tilde V_\mu- ig[\tilde V_\mu,\tilde
V_\nu],\nonumber\\ &&\tilde V_\mu=V_\mu-\frac{1}{3}\ Tr\ V_\mu\ ,\
\tilde v_\mu=v_\mu-\frac{1} {3}\ Tr\ v_\mu.
\end{eqnarray}
As it should be, ${\cal L}$ does not depend on $v$ and is therefore
invariant under $SU(3)_C$ transformations. For a check of local
$U(3)_P$ invariance we note
\begin{equation}\label{4.4b}
\delta v_\mu=i[\Theta_P,v_\mu]-\partial_\mu\Theta_P\ ,\ 
\delta a_\mu=i[\Theta_P,a_\mu]
\end{equation}
such that $v_\mu$ transforms the same as $-gV_\mu$ and $v_\mu+gV_\mu$
transforms homogeneously. One observes that the mass term for $\tilde
V_\mu$ appears in the $U(3)_P$-invariant combination $\chi_0^2\
Tr\{(\tilde v^\mu+g\tilde  V^\mu)(\tilde v_\mu+g\tilde V_\mu)\}$.  The
nonlinear field  $U$ only appears in derivative terms except for the
phase $\theta$ associated to the $\eta'$-meson. We associate $U$ in
the standard way with the pseudoscalar octet of Goldstone bosons
$\Pi^z$
\begin{equation}\label{4.5}
U=\exp(-\frac{i}{3}\theta)\exp\left(i\frac{\Pi^z\lambda_z}{f}\right
)={\exp}\left(-\frac{i}{3}\theta\right)\tilde U
\end{equation}
where the decay constant $f$ will be specified below.

The Lagrangian (\ref{4.4}) contains mass terms and kinetic terms for
an octet and a singlet of baryons. The same terms appear in the Higgs
picture of sect.~4. We want to use the physical baryon masses in order
to fix the Yukawa couplings $h$ and $\tilde h$. For this purpose we
need to identify the baryon singlet. We will argue that $\Lambda(1405)$,
which has the opposite parity as the 
nucleon or the lowest mass $\Lambda$-baryon, is an
interesting candidate. This requires that the ratio $\tilde
h\chi_0/h\sigma_0$ is negative. We observe that $h$ and $\tilde h$ can
be chosen (via the form of the Hubbard-Stratonovich transformation) to
be real and positive. Furthermore, we can use conventions where the
current quark masses $m_q$ are real and a parity conserving ground
state corresponds to positive $\sigma_0$. In the Higgs picture $C$ and
$P$ are conserved by real $\chi_0$. Similarly, this holds for the
expectation value of the gauge singlet $X_{kl,cd}$
(\ref{4.1}). However, the sign of $\chi_0$ is not fixed and on the
level of quarks we have no free phases left which could be used to
make $\chi_0$ always positive. The relative sign between the singlet
and octet contributions to the baryon masses $h\phi$ and $\tilde
h\chi_0$ is therefore observable. (Note that no $SU(3)_c$-gauge
transformation exists which reverses the sign of
$\chi_{ij,ab}$.) Consider now the case of a negative singlet mass term
$h\sigma_0+8\tilde h\chi_0/3\sqrt{6}$, i.e. negative $\chi_0$. On
the level of the nonlinear baryon fields we may apply a chiral phase
transformation acting only on the singlet field $N_1=Tr\ N/\sqrt3$
while leaving the octet field invariant. Choosing $\tilde
N_{1L}=iN_{1L},\ 
\tilde N_{1R}=-iN_{1R}$ we can
render the singlet mass positive, but the parity of the singlet is now
found opposite to the octet
\begin{equation}\label{PA}
P:\tilde N_{1L}\to\tilde N_{1R},\ \tilde N_{1R}\to-\tilde N_{1L}\quad,\quad
N_{8L}\to-N_{8R},\ N_{8R}\to N_{8L}.
\end{equation}
(In contrast, the $C$-parity is the same for the singlet and octet.)
Quite interestingly, there exists a singlet state $\Lambda(1405)$
with the opposite parity to the $\Lambda$ in the baryon octet! (A
singlet is not allowed in the ground state within the nonrelativistic
quark model.) We will associate the $\Lambda(1405)$ baryon with the
baryon singlet $S^0$
\begin{equation}\label{PB}
S^0\equiv\tilde N_1=i\gamma^5N_1=\frac{i}{\sqrt3}\gamma^5Tr\ N.
\end{equation}

From eq. (\ref{4.4}) one can now directly read the average (positive) 
masses  of the
baryon octet and singlet
\begin{equation}\label{4.6}
M_8=h\sigma_0-\frac{\tilde h}{3\sqrt6}\chi_0= 1.15\ {\rm GeV}\ ,\
M_1=-(h\sigma_0+\frac{8}{3}\frac{\tilde h}{\sqrt6}\chi_0) =1.4\ {\rm
GeV}.
\end{equation}
The singlet-octet mass splitting is proportional to the octet
condensate $\chi_0$
\begin{equation}\label{4.7}
M_1+M_8=-\frac{3}{\sqrt6}\tilde h\chi_0=2.55\ {\rm GeV}.
\end{equation}
With eq. (\ref{2.18a}) this determines the ratio
\begin{equation}\label{4.7a}
\frac{\tilde h}{g}=2.45.
\end{equation}
Similarly, one finds
\begin{equation}\label{4.7b}
h\sigma_0=\frac{1}{9}(8M_8-M_1)=866\ {\rm MeV}.
\end{equation}

\section{Mesons and their interactions}
\label{mesonsandtheir}
Before extracting the couplings between the physical mesons and the
baryons we have to understand the role of the local reparameterization
transformations $U(3)_P=SU(3)_P\times U(1)_P$.  First of all, we note
that $Tr\  v_\mu$ can be eliminated by $U(1)_P$-gauge transformations
and is therefore a pure gauge degree of freedom.  Similarly, $Tr\
V_\mu=-\frac{i}{g}\partial_\mu\ln\ \det\ v$ is also a gauge degree of
freedom \cite{CWSSB}. This explains why $Tr\  v_\mu$ and $Tr\  V_\mu$ only appear
in the term bilinear in the nucleon fields. The vector fields $\tilde
V_\mu$ are the gauge bosons of $SU(3)_P$. The nucleons transform as an
octet and a singlet under $SU(3)_P$ and the same  holds for the
bilinear $a_\mu$. The fields contained  in $W_L,W_R$ are antitriplets
and $U$ is a singlet.

For a discussion of the mesons we have to eliminate the redundance of
the local $U(3)_P$-transformations. We will employ the gauge choice
\begin{equation}\label{5.3a}
W_L^\dagger=W_R=\xi
\end{equation}
which implies
\begin{eqnarray}\label{5.4}
 &&U=\xi^2\ ,\ v_\mu=-\frac{i}{2}(\xi^\dagger \partial_\mu
\xi+\xi\partial_\mu\xi^\dagger),\nonumber\\
&&a_\mu=-\frac{i}{2}(\xi^\dagger\partial_\mu\xi-\xi\partial_\mu
\xi^\dagger)=-\frac{i}{2}\xi^\dagger\partial_\mu U\xi^\dagger,\nonumber\\
&&{\rm Tr}\ v_\mu=0\ ,\quad {\rm Tr}\ V_\mu=0\nonumber\\ &&\phi=\sigma_0U\ ,\
\xi_{ij,ab}=\frac{1}{\sqrt6}\chi_0(\xi_{ai}\xi_{jb}-
\frac{1}{3}U_{ab}\delta_{ij}),\\
&&(D_\mu\chi)_{ij,ab}=\frac{1}{\sqrt6}\chi_0
[\partial_\mu(\xi_{ai}\xi_{jb} -\frac{1}{3}U_{ab}\delta_{ij})
+ig((\xi\tilde V_\mu)_{ai}\xi_{jb}-\xi_{ai}(\tilde
V_\mu\xi)_{jb})].\nonumber
\end{eqnarray}
This gauge is singled out by the fact that $v_\mu$ contains no term
linear in $\Pi^z$ or $\theta$ \cite{CWSSB}
\begin{equation}\label{5.5}
v_\mu=\tilde v_\mu=-\frac{i}{2f^2}[\Pi,\partial_\mu\Pi]=
-\frac{i}{8f^2}[\lambda_y,\lambda_z]\Pi^y\partial_\mu
\Pi^z+...
\end{equation}
where we have chosen the parameterization
\begin{eqnarray}\label{X12}
&&\xi=\tilde\xi\exp\left(-\frac{i}{6}\theta\right)\ ,
\ det \tilde\xi=1\nonumber\\
&&\tilde\xi=\exp\left(\frac{i}{2f}\Pi^z\lambda_z\right)
=\exp\left(\frac{i}{f}\Pi\right).
\end{eqnarray}
As for any gauge where $v_\mu$ contains no term linear in $\Pi$ or
$\theta$, the term $\sim Tr\tilde v^\mu\tilde V_\mu$ contains only
interactions and does not affect the propagators. We can  therefore
associate $\Pi^z$ with the physical pions. Also the  term $Tr\tilde
v^\mu\tilde v_\mu$ involves at least four meson fields and is
irrelevant for the propagator of the pseudoscalar mesons.

The bilinear $a_\mu$ reads in the gauge (\ref{5.3a})
\begin{equation}\label{5.13}
a_\mu=\frac{1}{2f}\lambda_z\partial_\mu\Pi^z-\frac{1}{6}\partial_\mu\theta.
+O(\Pi^3,..)
\end{equation}
The effective action (\ref{4.4}) contains therefore a cubic coupling
between two baryons and a pion. As it should be for pseudo-Goldstone
bosons, this is a derivative coupling. For the gauge
(\ref{5.3a}) one has a simple realisation of the
discrete transformations $P$ and $C$ on the level of nonlinear 
fields\footnote{Note that in our matrix notation $C$ acts as $\psi_L\to
c\bar\psi_R^T,\ \psi_R\to-c\bar\psi^T_L$.}
\begin{eqnarray}\label{5.14}
P&:&N_L\to -N_R,\ N_R\to N_L,\nonumber\\ &&W_L\to W_R,\ W_R\to W_L,\
v\to v,\ U\to U^\dagger\ ,\
\xi\to\xi^\dagger\nonumber\\
&&v_\mu\to v_\mu,\ a_\mu\to -a_\mu,\ V_\mu\to V_\mu\nonumber\\
&&\nonumber\\ C&:& N_L\to c\bar N^T_R,\ N_R\to-c\bar N^T_L,\nonumber\\
&&W_L\to W_R^*,\ W_R\to W_L^*,\ v\to v^*,\ U\to U^T
\ ,\ \xi\to\xi^T\nonumber\\
&&v_\mu\to-v_\mu^T,\ a_\mu\to a_\mu^T,\ V_\mu\to -V_\mu^T.
\end{eqnarray}
The gauge condition (\ref{5.3a}) can also be written as
$W_LW_R=W_RW_L=1$ and is manifestly invariant under $P$ and $C$.  The
coupling ${\rm Tr}\bar N\gamma^\mu\gamma^5a_\mu N$ is a standard
coupling of baryons to the axial vector current.

The kinetic term for the pseudoscalar mesons can be inferred from
\begin{eqnarray}\label{5.6}
{\cal L}^{(P)}_{kin}&=&(\sigma_0^2+\frac{7}{36}\chi_0^2)\ Tr\{\partial
^\mu U^\dagger\partial_\mu U\}+\frac{1}{12}\chi^2_0\ \partial^\mu
\theta\ \partial_\mu\theta\nonumber\\
&=&\frac{2}{f^2}(\sigma_0^2+\frac{7}{36}\chi^2_0)\ \partial^\mu\Pi^z
\partial_\mu\Pi_z+\frac{1}{3}(\sigma_0^2+\frac{4}{9}\chi_0^2)\partial^\mu
\theta\partial_\mu\theta.
\end{eqnarray}
Its standard normalization fixes a combination of $\sigma_0$ and
$\chi_0$ in terms of the pseudoscalar decay constant $f$ which
corresponds to an average in the octet \cite{8}
$f=\frac{2}{3}f_K+\frac{1}{3}f_\pi=106$ MeV. For an optimal
quantitative estimate of the expectation values $\sigma_0, \chi_0$ the
so called ``partial Higgs effect'' should be included: the mixing
between the pseudoscalars contained in $U\ (\sim i\bar\psi\psi)$ and
those in the divergence of the axial-vector current
$(\sim\partial_\mu(\bar\psi\gamma^\mu\gamma^5\psi)$) introduces an
additional negative contribution (partial Higgs effect)
\begin{equation}\label{5.8a}
\Delta{\cal L}_{kin}^{(P)}=-\frac{\Delta^2_f}{4}
{\rm Tr}\{\partial^\mu \tilde U^\dagger \partial_\mu \tilde U\}
-\Delta^2_\theta \partial ^\mu\theta \partial_\mu\theta.
\end{equation}
One infers a standard renormalization of the kinetic term for
\begin{equation}\label{5.8b}
f^2=4\sigma^2_0+\frac{7}{9}\chi^2_0-\Delta^2_f=\kappa^2_f
(4\sigma_0^2+\frac{7}{9}\chi^2)
\end{equation}
with
\begin{equation}\label{5.8c}
\kappa_f=(1-\frac{\Delta_f^2}{4}(\sigma_0^2+\frac{7}
{36}\chi^2_0)^{-1})^{1/2}=(1+\frac{\Delta_f^2}{f^2})^{-1/2}\leq1.
\end{equation}
The deviation of $\kappa_f$ from one is one of the most important
effects of effective interactions not included in eq. (\ref{2.6}).
The effective interactions responsible for $\Delta^2_f$ also induce
$SU(3)$-violating contributions to the pseudoscalar wave function
renormalizations related to the strange quark mass\footnote{In the
notation of ref. \cite{8} one has
$\Delta^2_f/f^2=-X_\phi^-\bar\sigma_0^2Z_m^{-1}$.}. This has led to a
phenomenological estimate \cite{8}
\begin{equation}\label{5.8d}
\Delta^2_f\approx 0.45 f^2,\ \kappa_f\approx 0.83.
\end{equation}
The pseudoscalar decay constants provide now for a quantitative
estimate of the expectation values $\sigma_0$ and $\chi_0$
\begin{equation}\label{5.8}
f_0=2\sqrt{\sigma^2_0+\frac{7}{36}\chi^2_0}=f/\kappa_f=(\frac{2}{3}
f_K+\frac{1}{3}f_\pi)/\kappa_f=128\ {\rm MeV}.
\end{equation}
In the approximation (\ref{2.6}) one should use the leading order
relation $f=f_0$, keeping in mind that nonleading effects will
lower $f$ to
its physical value.

In the gauge (\ref{5.3a}) the nonlinear Lagrangian (\ref{4.4}) can now
be written in terms of the normalized physical fields $N, \tilde
V_\mu, \Pi$ and $\eta'$ as
\begin{equation}\label{5.15}
{\cal L}={\cal L}_N+{\cal L}_U+{\cal L}_V.
\end{equation}
The part involving baryons and the axial interactions with the
pseudoscalars reads
\begin{eqnarray}\label{5.16}
{\cal L}_N&=&i\ {\rm Tr}\{\bar N_8\gamma^\mu(\partial_\mu-i\gamma^5
\tilde a_\mu+\frac{i}{6H_{\eta'}}\gamma^5\partial_\mu\eta')N_8\}
+M_8 {\rm Tr}\{\bar N_8\gamma^5N_8\}\nonumber\\ &&+i\bar
N_1\gamma^\mu(\partial_\mu+\frac{i}{6H_{\eta'}}
\gamma^5\partial_\mu\eta')N_1-M_1\bar N_1\gamma^5N_1\nonumber\\
&&+\frac{1}{\sqrt3}\ {\rm Tr}\{\bar N_8\gamma^\mu\gamma^5\tilde
a_\mu\}N_1+
\frac{1}{\sqrt3}\bar N_1\gamma^\mu\gamma^5\ {\rm Tr}\{\tilde a_\mu N_8\}.
\end{eqnarray}
Here we use the octet and singlet baryon fields (recall
$N_1=-i\gamma^5S^0,\ \bar N_1=-i\bar S_0\gamma^5$)
\begin{equation}\label{84A}
N_1=\frac{1}{\sqrt3}{\rm Tr}\ N,\ N_8=N-\frac{1}{3}{\rm Tr}
N=N-\frac{1}{\sqrt3}N_1
\ ,\ 
{\rm Tr}\ N_8=0.
\end{equation}
The interaction with the pseudoscalars involves the axial vector
current
\begin{eqnarray}\label{5.17}
&&\tilde a_\mu=a_\mu-\frac{1}{3}\ {\rm Tr}\
a_\mu=\frac{1}{2f}\lambda_z
\partial_\mu \Pi^z+O(\Pi^3)\ ,\ {\rm Tr}\ \tilde a_\mu=0,\nonumber\\ 
&&a_\mu=
\tilde a_\mu-\frac{1}{6}\partial_\mu\theta=\tilde a_\mu-\frac{1}{6H_{\eta'}}
\eta'\quad,\quad\theta=\frac{1}{H_{\eta'}}\eta',
\end{eqnarray}
whereas the interaction with the vector current will be contained in
${\cal L}_V$. The kinetic term and interactions for the pseudoscalars
take the familiar form
\begin{equation}\label{5.18}
{\cal L}_U=\frac{f^2}{4}\ {\rm Tr}\{\partial^\mu\tilde U^\dagger
\partial_\mu\tilde U\}
+\frac{1}{2}\partial^\mu\eta'\partial_\mu\eta'-M^2_{\eta'}H^2_{\eta'}
\cos(\eta'/H_{\eta'})
\end{equation}
with
\begin{equation}\label{5.21}
\tilde U=\tilde\xi^2=
\exp\left(i\frac{\Pi^z\lambda_z}{f}\right),\ \det \tilde U=1.
\end{equation}
Finally, the terms involving vector mesons and vector currents are
grouped in
\begin{eqnarray}\label{5.22}
{\cal L}_V&=&\frac{1}{2}\ {\rm Tr}\{\tilde V^{\mu\nu}\tilde
V_{\mu\nu}\} +\bar M^2_\rho\ {\rm Tr}\{\tilde V^\mu\tilde
V_\mu\}+\frac{2}{g}\bar M^2_\rho {\rm Tr}\{\tilde V^\mu\tilde
v_\mu\}+\frac{1}{g^2}\bar M^2_\rho  {\rm Tr}\{\tilde v ^\mu\tilde
v_\mu\}\nonumber\\ &&-g \ {\rm Tr}\{\bar N_8\gamma^\mu N_8\tilde
V_\mu\}-\frac{g}{\sqrt3}( {\rm Tr}\{\bar N_8\tilde V_\mu\}\gamma^\mu
N_1+\bar N_1\gamma^\mu {\rm Tr}\{\tilde V_\mu N_8\})\nonumber\\
&&-{\rm Tr}\{\bar N_8\gamma^\mu\tilde v_\mu N_8\}-\frac{1}{\sqrt3}
({\rm Tr}\{\bar N_8\tilde v_\mu\}\gamma^\mu N_1+\bar N_1\gamma^\mu
{\rm Tr}\{\tilde v_\mu N_8\}).
\end{eqnarray}

In the effective action (\ref{5.15}) we have replaced the free
parameters by $M_8, M_1$, $\bar M_\rho, M_{\eta'} , f$ and $g$. We
note that the number of parameters is reduced to six since only one
particular combination of $\nu$ and $\nu'$ appears in $M_{\eta'}$. 
At this stage of our investigation only the effective gauge coupling
has not yet been determined.
The proportionality constant $H_{\eta'}$ appearing in the couplings of
$\eta'$ reads \cite{CWSSB}
\begin{equation}\label{88A}
H_{\eta'}=\left(\frac{1+\frac{16}{7}x}{1+x}\right)^{1/2}\frac{\kappa_\theta}{\kappa_f}
\frac{(2f_K+f_\pi)}{3\sqrt{6}}
\end{equation}
where $\kappa_\theta/\kappa_f$ accounts for the details of the
``partial Higgs effect''.  It can be related to the decay width of
$\eta'$ into two photons \cite{8}. 

The interactions between individual  baryons
and mesons can be extracted by inserting the explicit representations
\begin{eqnarray}\label{5.23}
&&\Pi=\frac{1}{2}\Pi^z\lambda_z=\frac{1}{2}\left(
\begin{array}{ccc}
\pi^0+\frac{1}{\sqrt3}\eta\ ,&\ \sqrt2 \pi^+\ ,&\ \sqrt 2K^+\\
\sqrt2\pi^-\ ,&\ -\pi^0+\frac{1}{\sqrt3}\eta\ &\ \sqrt2 K^0\\
\sqrt2 K^-\ ,& \ \sqrt2\bar K^0\ ,&\ -\frac{2}{\sqrt3}\eta\end{array}
\right)\nonumber\\
&&\nonumber\\ &&\quad \tilde V_\mu=\frac{1}{2}\tilde
V_\mu^z\lambda_z=\frac{1}{2}
\left(\begin{array}{ccc}
\rho^0_\mu+\frac{1}{\sqrt3}V^8_\mu\ ,&\ \sqrt2 \rho^+_\mu\ ,&\ \sqrt2K^{*+}_\mu\\
\sqrt2\rho^-_\mu\ ,&\ -\rho^0_\mu+\frac{1}{\sqrt3}V^8_\mu\ &\ \sqrt2 K^{*0}_\mu\\
\sqrt2 K^{*-}_\mu\ ,& \ \sqrt2\bar K^{*0}_\mu,&-\frac{2}{\sqrt3}V^8_\mu \end{array}
\right)\nonumber\\
&&\nonumber\\ &&N_8=\left(
\begin{array}{ccc}
\frac{1}{\sqrt2}\Sigma^0+\frac{1}{\sqrt6}\Lambda^0\ ,&\ \Sigma^+\ ,&\ p\\
\Sigma^-\ ,&\ -\frac{1}{\sqrt2}\Sigma^0+\frac{1}{\sqrt6}\Lambda^0\ ,& \ n\\
\Xi^-\ ,&\ \Xi^0\ ,&\ -\frac{2}{\sqrt6}\Lambda^0 \end{array}\right)
\nonumber\\
&&\nonumber\\ &&\bar N_8=\left(
\begin{array}{ccc}
\frac{1}{\sqrt2}\bar\Sigma^0+\frac{1}{\sqrt6}\bar\Lambda^0\ .&
\ \bar\Sigma^-\ ,&\ \bar\Xi^-\\ 
\bar \Sigma^+\ ,&\ -\frac{1}{\sqrt2}\bar\Sigma^0+\frac{1}{\sqrt6}
\bar\Lambda^0\ ,&\ \bar\Xi^0\\
\bar p\ ,&\ \bar n\ ,& \ -\frac{2}{\sqrt6}\bar\Lambda^0\end{array}
\right)
\end{eqnarray}
It is obvious that the effective action (\ref{5.15}) predicts a
multitude of different interactions between the low mass hadrons. The
electromagnetic interactions are incorporated by covariant derivatives.

Finally, the addition of the explicit chiral symmetry breaking by the
current quark masses (\ref{2.8})
\begin{equation}\label{5.28}
{\cal L}_j=-\frac{1}{2}Z_\phi^{-1/2}a_q\sigma_0\ {\rm Tr}\{\bar
m(U+U^\dagger)\}
\end{equation}
leads to mass terms for the pseudoscalars \cite{Mass} and contributes
to their interactions, in accordance with chiral perturbation theory
(cf. eq. (\ref{A24})).

\section{Electromagnetic interactions}
\label{electromagneticinteractions}
As a first probe of our picture we may use the electromagnetic
interactions of mesons and baryons. The electromagnetic interactions
of the mesons are all contained in the covariant kinetic term for the
scalars
\begin{eqnarray}\label{E1}
(D^\mu\gamma_{ij,ab})^*D_\mu\gamma_{ij,ab}&=&
\chi^2_0\ {\rm Tr} \{(\hat v^\mu+g\tilde V^\mu-\tilde e
\tilde B^\mu\tilde Q)(\hat v_\mu+g\tilde V_\mu-\tilde e\tilde B_\mu\tilde
Q)\}\nonumber\\ &&+\frac{f^2}{4}\ {\rm Tr}\{(D^\mu U)^\dagger D_\mu
U\}+\frac{1}{12}
\chi^2_0\ \partial^\mu\theta \partial_\mu\theta
\end{eqnarray}
with generator $\tilde{Q}=diag(2/3,-1/3,-1/3)$ for the electric charge.
Here $D_\mu U=\partial_\mu U-i\tilde e\tilde B_\mu[\tilde Q,U]$ and
the covariant vector current reads (with $\tilde
D_\mu\xi=\partial_\mu\xi-i\tilde e\tilde B_\mu[\tilde Q,\xi])$
\begin{eqnarray}\label{E2}
\hat v_\mu&=&-\frac{i}{2}(W_L^\dagger\tilde D_\mu W_L+W_R^\dagger
\tilde D_\mu W_R)\nonumber\\
&=&-\frac{i}{2}(\xi^\dagger\tilde D_\mu\xi+\xi\tilde
D_\mu\xi^\dagger)\nonumber\\ &=&v_\mu-\frac{\tilde e}{2}\tilde
B_\mu(\xi^\dagger\tilde Q\xi+\xi
\tilde Q\xi^
\dagger-2\tilde Q).
\end{eqnarray}
Observing that both $\hat v_\mu$ and $g\tilde V_\mu-\tilde e\tilde
B_\mu\tilde Q$ transform homogeneously with respect to the
electromagnetic gauge transformations
\begin{equation}\label{E2a}
\delta_{em}\hat v_\mu=i\beta[\tilde Q,\hat v_\mu],\ \delta_{em}
(g\tilde V_\mu-\tilde e\tilde B_\mu\tilde Q) =i\beta[Q,(g\tilde
V_\mu-\tilde e\tilde B_\mu\tilde Q)]
\end{equation}
the gauge invariance of (\ref{E1}) can be easily checked. The
particular combination of vector currents in (\ref{E1}) is dictated by
the combination of electromagnetic gauge invariance and local
reparameterization symmetry.  Only this combination transforms
homogeneously with respect to both $U(3)_P$ local reparameterizations
and electromagnetic $U(1)$ gauge transformations. 

The interactions (\ref{E1}) coincide with those of a ``hidden local
chiral symmetry'' \cite{Bando} if one replaces $\hat v_\mu$ by
$v_\mu$.  (The small difference between our result and the ``hidden
symmetry''  approach is due to the somewhat different electromagnetic
transformation properties of the nonlinear fields). If one restricts
the discussion to $\rho$-mesons and pions, $\tilde
V_\mu=\frac{1}{2}\vec\rho_{V\mu}\vec\tau,
\Pi=\frac{1}{2}\vec\pi\vec\tau$, and neglects the difference between
$\tilde e$ and $e$, one finds for the term involving the vector mesons
and currents
\begin{eqnarray}\label{E3}
{\cal L}_{VV}&=&af^2_\pi\ {\rm Tr}(\hat
v_\mu+\frac{1}{2}g_\rho\vec\rho_{V\mu}
\vec\tau-\frac{1}{2}e\tilde B_\mu\tau_3)^2\nonumber\\
&=&af^2_\pi\ {\rm Tr}\left[
\frac{1}{4f_\pi^2}((\vec\pi\times \partial_\mu\vec\pi)\vec\tau)+\frac{1}{2}g_\rho
(\vec\rho_{V\mu}\vec\tau)-\frac{1}{2}e\tilde B_\mu\tau_3
\right.\nonumber\\
&&\left.-\frac{e}{4f^2_\pi}\tilde
B_\mu(\pi_3(\vec\pi\vec\tau)-(\vec\pi\vec\pi)\tau_3)
\right]^2+...
\end{eqnarray}
For the second equality in eq. (\ref{E3}) we have only retained terms
quadratic in $\pi$ in an expansion of $\hat v_\mu$ and we have
replaced $f$ by $f_\pi$.  The first expression in eq. (\ref{E3}) is
actually more general than the result of the particular effective
action (\ref{2.6}) for which one has
\begin{eqnarray}\label{E4}
g_\rho&=&g,\nonumber\\ a&=&\frac{\chi_0^2}{f_\pi^2}=\frac{9}
{7\kappa_f^2}\frac{x}{1+x}\frac{f^2}{f^2_\pi}\approx 2.4\frac{x}{1+x}.
\end{eqnarray}
In consequence of the symmetries, additional interactions will  change
the values of a $a$ and $g_\rho$ without affecting the structure of
the invariant.  Such additional interactions will also generate new
invariants involving higher derivatives \cite{CWSSB}.

The contributions in eq (\ref{E3}) with canonical dimension $\leq4$
can be written in the form
\begin{eqnarray}\label{E5}
{\cal L}_{VV}&=&\frac{1}{2}M_\rho^2\vec\rho^\mu_V\vec\rho_{V\mu}-eg_
{\rho\gamma}\rho^\mu_{V3}\tilde B_\mu+\frac{1}{2}m_B^2\tilde
B^\mu\tilde B_\mu\nonumber\\
&&+g_{\rho\pi\pi}\vec\rho_V^\mu(\vec\pi\times \partial_\mu\vec\pi)
+g^{(V)}_{\gamma\pi\pi}\tilde B^\mu (\vec\pi\times
\partial_\mu\vec\pi)_3\nonumber\\ &&+g_{\rho\gamma\pi\pi}\tilde
B_\mu[(\vec\pi\vec\pi)\rho^\mu
_{V3}-(\vec\rho^\mu_V\vec\pi)\pi_3]\nonumber\\
&&-\frac{1}{2}ae^2\tilde B_\mu\tilde B^\mu[(\vec\pi\vec\pi-\pi^2_3]+...
\end{eqnarray}
with
\begin{eqnarray}\label{E6}
M^2_\rho&=&ag^2_\rho f_\pi^2\ ,\ g_{\rho\gamma}=ag_\rho f_\pi^2\ ,\
g_{\rho\pi\pi}=\frac{1}{2}ag_\rho,\nonumber\\ m_B^2&=&ae^2f_\pi^2\ ,\
g^{(V)}_{\gamma\pi\pi} =-\frac{1}{2}ae\ ,\
g_{\rho\gamma\pi\pi}=\frac{1}{2}aeg_\rho.
\end{eqnarray}
The effective action (\ref{2.6}) therefore leads to
 the very successful KSFR relation \cite{KSFR}, \cite{Bando}
\begin{equation}\label{E7}
g_{\rho\gamma}=2f_\pi^2g_{\rho\pi\pi}
\end{equation}
which relates the decay $\rho\to 2\pi$ (with $g_{\rho\pi\pi}
\approx 6$, see next section) to the electromagnetic properties
of the $\rho$-meson, in particular the decay $\rho_0\to e^+e^-$ (with
$\Gamma(\rho_0\to e^+e^-)=6.62$ keV and $g_{\rho\gamma}=0.12$ GeV$^2$).

The relation
\begin{equation}\label{E8}
M_\rho^2=\frac{4}{a}g^2_{\rho\pi\pi}f^2_\pi
\end{equation}
requires $a=2.1$, implying
\begin{equation}\label{7.10a}
\chi_0=135\ {\rm MeV}.
\end{equation}
One could use eq. (\ref{E4}) for an estimate of the relative octet
contribution to the pion decay constant and eq. (\ref{2.18a}) 
for an estimate of $g$
\begin{equation}\label{7.10b}
x=7\quad, \quad \sigma_0=22\ {\rm MeV}\quad,\quad g=6.3.
\end{equation}
This should, however, not be taken too literally in view of possible
substantial $SU(3)_V$-violating effects from the nonzero strange quark
mass. Also higher order operators may affect the relation
(\ref{E4}). Furthermore, one should include the corrections from
additional invariants. However, the possible
 additional  invariants only affect the
cubic and higher vertices, but not the mass terms. In particular, the
relation
\begin{equation}\label{7.10d}
g_{\rho\gamma}=g\chi^2_0=\frac{\bar M^2_\rho}{g}
\end{equation}
will only be modified by $SU(3)$-violating effects. It can be used for
an independent estimate of $g$, yielding $g=6$. Because of the
relative robustness of this last estimate we will use the observed
value of $g_{\rho\gamma}$ as an input and assume $g=6$. All free
parameters are then fixed by observation and the decay rate for
$\rho\to 2\pi$ is a first ``prediction'' of our model.

The electromagnetic $\gamma\pi\pi$ and $\gamma\gamma\pi\pi$ vertices
also receive contributions from
\begin{eqnarray}\label{E9}
\frac{1}{4}f^2_\pi{\rm Tr}\{D^\mu U^\dagger D_\mu U\}&=&\frac{1}{2}
\partial^\mu\vec\pi\partial_\mu\vec\pi\\
&&+e\tilde B^\mu(\vec\pi\times \partial_\mu\vec\pi)_3+\frac{1}{2}
e^2\tilde B_\mu\tilde B^\mu(\vec\pi\vec\pi-\pi^2_3)+...\nonumber
\end{eqnarray}
One sees that for $a=2$ the two contributions (\ref{E5}) and
(\ref{E9}) to the direct $\gamma\pi\pi$ vertex cancel. The
electromagnetic interactions of the pions are then dominated by
$\rho$-exchange (vector dominance), in agreement with
observation. Otherwise stated, our model leads for the direct
$\gamma\pi\pi$-coupling  to the realistic relation
\begin{equation}\label{E10}
g_{\gamma\pi\pi}
=e\left(1-\frac{2g^2_{\rho\pi\pi}f^2_\pi}{M_\rho^2}\right).
\end{equation}
The vertex $\sim g_{\rho\gamma\pi\pi}$ contributes\footnote{This
vertex is absent in ref. \cite{Bando}.} to rare decays like
$\rho_0\to\pi^+\pi^-\gamma$, with
\begin{equation}\label{E11}
g_{\rho\gamma\pi\pi}=eg_{\rho\pi\pi}.
\end{equation}

We conclude that the electromagnetic interactions of the pseudoscalars
as well as the vector mesons can be considered as a successful test of
our simple model. The appearance of a local nonlinear
reparameterization symmetry is a direct consequence of the
``spontaneous breaking'' of color. This symmetry, combined with the
simple effective action  (\ref{2.6}), has led to the KSFR relation
(\ref{E7}) and to vector dominance (\ref{E10}). At this stage the
relations of the above discussion should be taken with a 20-30 percent
uncertainty. In particular, the $SU(3)$-violation due to the nonzero
strange quark mass needs to be dealt with more
carefully. Nevertheless, the $\rho\to 2\pi$ decay and the
electromagnetic decays are all consistent and have allowed us a first
determination of the size  of the octet condensate $\chi_0$. The octet
condensate is larger than the singlet condensate and dominates the
pseudoscalar decay constant $f$.

Finally, the connection between the above  discussion and the Higgs picture
developed in sect. \ref{higgspicture} is easily established if one
realizes that the  field
\begin{eqnarray}\label{7.7}
G_\mu&=&-\sqrt3\ Tr\ \{\tilde Q\tilde V_\mu\} =\sqrt3\ Tr\ \{\tilde
Q(vA_\mu^T v^\dagger+\frac{i}{g}
\partial_\mu vv^\dagger)\}\nonumber\\
&=&-\left(\frac{\sqrt3}{2}\rho^0_\mu+\frac{1}{2}V^8_\mu\right)
\end{eqnarray}
is the nonlinear correspondence of $\tilde G_\mu$ in
sect. \ref{higgspicture}.  With respect to the electromagnetic gauge
transformations it transforms inhomogeneously
\begin{equation}\label{7.8}
\delta_{em} G_\mu=-\frac{2}{\sqrt3 g}\partial_\mu\beta=-\frac{1}{\tilde
g}\partial_\mu\beta
\end{equation}
such that the linear combinations (\ref{3.7}) $R_\mu=
\cos\theta_{em} G_\mu+\sin\theta_{em}\tilde B_\mu,B_\mu=\cos
\theta_{em}\tilde B_\mu-\sin\theta_{em}G_\mu$ have the 
transformation properties of a heavy neutral boson and a photon
\begin{equation}\label{7.9}
\delta_{em}B_\mu=\frac{1}{e}\partial_\mu\beta\quad,\quad\delta_{em}
R_\mu=0.
\end{equation}
Inserting in ${\cal L}_V$ (\ref{5.22}) $\tilde V_\mu=
-\frac{\sqrt3}{2}G_\mu\tilde Q,\tilde v_\mu=0$ and adding terms from
covariant derivatives (\ref{3.1}) involving $\tilde B_\mu$
(cf. (\ref{E3})), we recover ${\cal L}^{(2)}_{em}$ (\ref{3.6}) with
$\tilde G_\mu$ replaced by $G_\mu$. The   (extended) electromagnetic
interactions of the baryon octet\footnote{We omit here terms involving
the baryon singlet $N_1$ and take $\Pi=0$.} read
\begin{eqnarray}\label{7.10}
{\cal L}_{VN,0}&=&eB_\mu\ Tr\ \{\bar N_8\gamma^\mu[\tilde
Q,N_8]\}\nonumber\\ &&+\tilde g\cos\theta_{em}R_\mu\ Tr\ \{\bar
N_8\gamma^\mu (N_8\tilde Q+tg ^2\theta_{em}\tilde QN_8)\}.
\end{eqnarray}
One finds the standard coupling between the photon $B_\mu$ and the
baryons according to their integer electric charge. More generally, we
conclude from
\begin{eqnarray}\label{7.11}
&&\tilde e\tilde B_\mu=e(B_\mu+tg\theta_{em}R_\mu)\nonumber\\ &&\tilde
g G_\mu=\tilde g\cos\theta_{em}R_\mu-eB_\mu
\end{eqnarray}
that  the charged leptons have a small direct coupling to a linear
combination of the $\rho^0, \omega$ vector mesons corresponding to the
term $e\  tg\theta_{em}R_\mu$ in  $\tilde e\tilde B_\mu$. Similarly,
the photon has a hadronic coupling from the term $-eB_\mu$ in $\tilde
g G_\mu$.  These mixing effects are governed by the coupling
$g_{\gamma\rho}$ (\ref{E7}).

\section{Vector mesons}
\label{vectormesons}
The vector mesons acquire a mass through the Higgs mechanism.  They
are also unstable due to the decay into two pseudoscalar mesons. Their
interactions are contained in ${\cal L}_V$ (\ref{5.22}).  In
particular, the cubic vertex between one vector meson and two
pseudoscalars is
\begin{eqnarray}\label{7.1}
{\cal L}_{V\pi\pi}&=&2g\chi_0^2\ {\rm Tr}\{\tilde V^\mu\tilde v_\mu\}
=-\frac{ig\chi^2_0}{f^2}\ {\rm Tr}\{[\Pi,\partial_\mu\Pi]\tilde V^\mu\}
\nonumber\\
&=&-2i g_{\rho\pi\pi}\ Tr\{[\Pi,\partial_\mu\Pi]\tilde V^\mu\}
\end{eqnarray}
with
\begin{equation}\label{7.2}
g_{\rho\pi\pi}=\frac{g}{2}\frac{\chi^2_0}{f^2}=\frac{M^2_\rho}{2gf^2}.
\end{equation}
Considering only the effective action (\ref{2.6}), one has
$\kappa_f=1$ and would infer
\begin{equation}\label{7.2a}
g_{\rho\pi\pi}=
\frac{9x}{14(1+x)}g\approx 4.6\left(\frac{x}{1+x}\right)^{1/2}.
\end{equation}
If we restrict these interactions to the $\rho$-mesons $\tilde
V_\mu=\frac{1}{2}\vec\rho_{V\mu}\vec\tau$ and  pions 
$\Pi$$=$$\frac{1}{2}\vec\pi\vec\tau, v_\mu=\frac{1}{4f^2}(
\vec\pi\times\partial_\mu\vec\pi)\vec\tau$, we obtain
the familiar form
\begin{equation}\label{7.2b}
{\cal
L}_{\rho\pi\pi}=\frac{g\chi_0^2}{2f^2}\epsilon_{ijk}\pi^i\partial_\mu
\pi^j\rho_V^{k\mu}
=g_{\rho\pi\pi}(\vec\pi\times\partial_\mu\vec\pi)\vec\rho _V^\mu.
\end{equation}
It is straightforward to compute the decay rate $\rho\to 2\pi$ as
\begin{equation}\label{7.3}
\Gamma(\rho\to\pi\pi)=\frac{g^2_{\rho\pi\pi}}{48\pi}
\frac{(M_\rho^2-4M^2_\pi)^{3/2}}{M_\rho^2}\simeq
150\ {\rm MeV}
\end{equation}
and one infers the phenomenological value $g_{\rho\pi\pi}\simeq
6.0$. The discrepancy with eq. (\ref{7.2a}) reflects the difference
between the realistic value $a\approx 2$ and a value of $a$ which
would result from (\ref{E4}) for $\kappa_f=1,\ f_\pi=f$. The estimate
(\ref{7.2a}) receives, however, important corrections.  Omitted
$SU(3)_V$-violating effects from the strange quark mass result in
corrections $\sim$ 30 \%. Furthermore, since eq. (\ref{E4}) is no
symmetry relation, it is subject to modifications from the inclusion
of additional fields. It is precisely the inclusion of these effects
  in eq. (\ref{E4}) (i.e. $f_\pi\not= f,\ \kappa_f<1$) that makes the
prediction  of our model for $g_{\rho\pi\pi}$ agree with observation.

The effective coupling between $\rho$-mesons and baryons (\ref{5.22})
obeys
\begin{equation}\label{7.6}
{\cal L}_{\bar NN_\rho}=-\frac{g}{2}\ {\rm Tr}\ \{\bar N_8\gamma^\mu
N_8\vec\tau\}\vec\rho_{V\mu}
\end{equation}
where we omit from now on electromagnetic effects.  This implies that
the effective action (\ref{2.6}) does not contain a direct coupling of
the $\rho$-mesons to protons and neutrons.  The $\rho$-mesons only
couple to strange baryons why the nucleon-nucleon interactions in the
vector channel are mediated by the exchange of the $\omega$-meson.
In the approximation of the effective
action (\ref{2.6}) possible contributions to the nucleon-nucleon
interactions in the isospin triplet vector channel could only arise
through two-pion interactions $\sim {\rm Tr}\ \{\bar
N_8\gamma^\mu\tilde v_\mu N_8\}$.  

More generally, one may compute in
our model  the
effective nucleon-nucleon interactions 
and  compare them with experimental information. The exchange of the
mesons generates effective vertices for nucleon-nucleon scattering in
the form of terms $\sim(\bar N N)^2$. For their determination
one should solve the field
equations for $\Pi$ and $\tilde V_\mu$ as functionals of the baryon
fields $N_8$ in bilinear order $\sim
\bar N_8N_8$. The solution has to be reinserted into the effective
action. As a result of this procedure one finds nucleon-nucleon
interactions in the pseudoscalar channel and in the isospin-singlet
vector channel mediated by the exchange of $V_{8\mu}$ but not in the
isospin-triplet vector channel. (The-$\sigma$ exchange term in the
scalar channel is also contained in our model once the non-Goldstone
scalar excitations in $\phi$ and $\chi$ are included.) This absence of
isospin-triplet vector channel nucleon-nucleon interactions seems not
to be consistent with observation \cite{9} but the size of the
relevant coupling remains debated.

In summary, three shortcomings indicate that the effective action
(\ref{2.6}) gives only an insufficient picture of the hadronic
interactions in the vector channel: (i) the absence of a physical
$SU(3)$-singlet vector state (the ninth vector meson), (ii) the
inaccurate estimate of the parameter $a$ in eq. (\ref{E4}) for
$\kappa_f=1$, (iii) the absence of nucleon-nucleon interactions in the
isospin-triplet vector channel. In addition, no axial-vector mesons
are present. These shortcomings can be overcome once we include the
effective four-quark interactions in the color singlet vector and
axial-vector channel. The successful  relations (\ref{E7}),
(\ref{E10})  and (\ref{E11}) can be maintained, whereas  the parameter
$a$  and the relation  between $M_\rho, g_{\rho\pi\pi}$ and $f_\pi$
will be modified. Mixing effects with the divergence of the axial
vector induce the correction (\ref{5.8a}) and therefore
$\kappa_f<1$. The effects from vector and axial-vector four-quark
interactions are discussed in detail in \cite{CWSSB}. Besides the
``partial Higgs effect'' (\ref{5.8a}) they account for the missing
nucleon-nucleon interactions in the isospin triplet channel. They also
contain the missing ninth vector meson, i.e. the $SU(3)_C$-singlet
state, as well as the axial-vector mesons. Unfortunately, the
additional effective interactions are parameterized by new unknown
couplings.

At this point we also can make the connection to the formulation in
sect. 2 which uses only gauge invariant composite operators. Indeed,
we may introduce additional fields for the gauge singlet fermion
bilinears transforming as a vector
\begin{equation}\label{B1}
\rho^\mu_L\sim\bar{\psi}_L\gamma^\mu\psi_L~,~\rho^\mu_R\sim\bar{\psi}_R
\gamma^\mu\psi_R.
\end{equation}
Neglecting the axial vector degrees of freedom this will lead to an
effective action\footnote{See ref. \cite{CWSSB}, appendix A for a
detailed description of the structure of the effective action in
presence of $\rho^\mu_L,\rho^\mu_R$, including a discussion of axial 
vectors.} $\Gamma[A_\mu,\varphi,\chi,\rho]$
which depends in addition on the fields
\begin{equation}\label{122A}
\rho^\mu=\frac{1}{\sqrt{2}}(\rho^\mu_L+\rho^\mu_R).
\end{equation}
In presence of the ``octet condensate'' the fields $\rho^\mu$ and
$A^\mu$ (more precisely $\rho^\mu$ and $V^\mu$) will mix and the low
mass physical vector mesons correspond to an appropriate linear
combination. One may reduce the degrees of freedom in the effective
action by solving the field equations for one of the vector fields in
presence of the other and reinserting the result into the effective
action. After elimination of $\rho^\mu$ in favor of $A^\mu$ the 
effective action (\ref{2.6}) may be viewed as the lowest order in a derivative
expansion for the octet vector mesons. The missing couplings and the
singlet vector meson are provided by $\rho^\mu$ \cite{CWSSB}. In
contrast, the formulation of sect. \ref{functionalintegral} in terms
of gauge invariant vector fields is recovered if we eliminate $A_\mu$
and $\chi$ in favor of $\rho^\mu,\varphi$,
i.e. $\Gamma[A_\mu,\varphi,\chi,\rho]\rightarrow\Gamma[\varphi,\rho]$. The
result is then precisely the effective action for the gauge invariant
bilinears similar to sect. \ref{functionalintegral}. From there
 we can extract directly the gauge invariant
correlation functions as ``measured'', for example, on the lattice.

More in detail, the mixing between $\rho^\mu$ and $A^\mu$ results from
invariants of the type \cite{CWSSB}
\begin{equation}\label{B2}
(D_\mu\chi)^*_{ijab}\chi_{ijac}
(\rho^\mu_L)_{cb}+\cdots\Rightarrow\chi^2_0\rho^\mu A_\mu.
\end{equation}
Expanding the effective action in terms quadratic in $A^\mu$ and
$\rho^\mu$ one obtains
\begin{equation}\label{B3}
{\cal L}_{\rho A}=\frac{1}{2}M^2\rho^\mu\rho_\mu +\frac{1}{2}A^\mu
(-\partial^2+g^2\chi^2_0)A_\mu-\beta\chi^2_0\rho^\mu A_\mu+\dots
\end{equation}
This leads to a lowest order field equation for $A^\mu$
\begin{equation}\label{B4}
(-\partial^2+g^2\chi^2_0)A^\mu=\beta\chi^2_0\rho^\mu.
\end{equation}
Reinserting the solution into ${\cal L}_{\rho A}$ (\ref{B3}) we find
the term quadratic in $\rho^\mu$
\begin{equation}\label{B5}
{\cal L}_\rho=\frac{1}{2}\rho^\mu\frac{M^2}{-\partial^2+g^2\chi^2_0}
(-\partial^2+g^2\chi^2_0-\beta^2\chi^4_0/M^2)\rho_\mu.
\end{equation}
It is now straightforward to extract the $\rho$-meson mass from the
field equation for $\rho^\mu$
\begin{equation}\label{B6}
(-\partial^2+\bar{m}^2_\rho)\rho^\mu=0.
\end{equation}
We see that the effective mass $\bar{m}_\rho$ is essentially the one
found from the effective action (\ref{2.6}), with a correction
$\sim\beta^2$ due to the mixing effect
\begin{equation}\label{B7}
\bar{m}^2_\rho=g^2\chi^2_0\left(1-\frac{\beta^2\chi^2_0}{g^2M^2}\right).
\end{equation}
In summary, the effective action in presence of the scalar color octet
field $\chi$ can also be considered as an intermediate step for the
computation of the action for
gauge singlet $\bar{q}q$ bilinears.

\section{Interactions of pseudoscalar mesons}
\label{interactionsof}
The cubic interactions between the pseudoscalar and the baryon octets
are usually parameterized by
\begin{equation}\label{99.1}
{\cal L}_N^{(p)}=F\ {\rm Tr}\{\bar N_8\gamma^\mu \gamma^5 [\tilde
a_\mu,N_8]\}+D\ {\rm Tr}\{\bar N_8\gamma^\mu\gamma^5\{\tilde a_\mu,
N_8\}\}.
\end{equation}
Experimental values are $F=0.459\pm0.008$ and $D=0.798\pm0.008$. The
effective action (\ref{2.6}) leads to (cf. eq. (\ref{5.16}))
$F=D=0.5$. This can be considered as a good achievement since 
these couplings remain completely undetermined by the chiral
symmetry. Also the
reparameterization symmetry does not restrict the coupling constants
$F$ and $D$. (This contrasts with the coupling of the vector current
$\tilde v_\mu$.)  In our context the relation $F=D=0.5$ is directly
connected to the origin of these couplings from the quark kinetic term
and therefore to quark-baryon duality. The
partial Higgs effect provides for a correction $F+D=1+y_N$ and
one infers $y_N\approx 0.26$ \cite{CWSSB}. Contributions to $D-F\approx0.34$ have
to be generated from other higher-order invariants, as, for example, a
momentum-dependent Yukawa coupling involving $D_\mu\chi$. We note that
$D-F$ contributes to the  $\eta$-nucleon coupling but not to the
interaction between protons, neutrons and pions. The latter can be
written in a more conventional form with the nucleon doublet ${\cal
N}^T=(p,n)$ and $\tilde a_\mu$ restricted to a $2\times 2$ matrix (by
omitting the last line and column)
\begin{eqnarray}\label{9.1a}
{\cal L}^{(\pi)}_{\cal N}&=&g_A\bar{\cal N}\gamma^\mu\gamma^5
\tilde a_\mu{\cal N}\nonumber\\
g_A&=&F+D.
\end{eqnarray}
The inclusion of weak interactions will replace the derivative  in the
definition (\ref{5.4}) of $\tilde a_\mu$ by a covariant derivative
involving a coupling to the $W$-boson \cite{CWSSB}.  The constant
$g_A$ will therefore appear in the $\beta$-decay rate of the neutron.

The vertex responsible for the decay of the baryon singlet
$S^0\to\Sigma\pi$ involves no $\gamma^5$ as appropriate in our
conventions for a baryon with opposite parity
\begin{eqnarray}\label{131A}
{\cal L}_{S{\Sigma}\pi} 
&=& -\frac{i}{\sqrt3}{\rm Tr}\{\bar N_8\tilde
a_\mu\}\gamma^\mu S^0+h.c.\nonumber\\
&=& -\frac{i}{\sqrt6f_\pi}(\bar\Sigma^+\partial_\mu\pi^+
+\bar\Sigma^-\partial_\mu\pi^-
+\bar\Sigma^0\partial_\mu\pi^0)\gamma^\mu S^0+h.c.
\end{eqnarray}
It involves no additional parameter such that the  decay width is computable.

For a discussion of the self-interactions of the pseudoscalar mesons
we first expand eq. (\ref{5.18}) in powers of
$\Pi=\frac{1}{2}\Pi^z\lambda_z$
\begin{eqnarray}\label{99.2}
{\cal L}^{(\pi)}&=&\frac{f^2}{4}\ Tr\{\partial^\mu\exp(-\frac{2i}{f}
\Pi)\partial_\mu\exp(\frac{2i}{f}\Pi)\}\\
&=&Tr\{\partial^\mu\Pi\partial_\mu\Pi\}+\frac{1}{f^2}\ Tr\{
\partial^\mu\Pi^2\partial_\mu\Pi^2-\frac{4}{3}\partial^\mu
\Pi\partial_\mu\Pi^3\}+...\nonumber
\end{eqnarray}
Similarly, the current quark mass term (\ref{5.28}) contributes
\begin{eqnarray}\label{9.3M}
{\cal L}_j&=&{\rm Tr}\left\{ M^2_{(p)}\Pi^2\right\}-{\rm Tr}
\left\{\frac{M^2_{(p)}}{3f^2}\Pi^4\right\}+...\nonumber\\
M^2_{(p)}&=&2Z^{-1/2}_\phi a_q\sigma_0f^{-2}\bar m.
\end{eqnarray}
This yields the low-momentum four-pion interactions.  Further
effective  interactions arise from the exchange of vector mesons
according to eq. (\ref{5.22}).  They are obtained by substituting in
${\cal L}$ for the vector meson fields $\tilde{V}_\mu$ the solution of
the field equation in  presence of pseudoscalars
\begin{equation}\label{99.3}
\tilde V_\mu=-\frac{\bar M^2_\rho}{g}G^{(\rho)\nu}_\mu\tilde v_\nu+...
\end{equation}
Here the vector meson propagator $G^{(\rho)}$ obeys
\begin{equation}\label{99.4}
G^{(\rho)\ \nu}_{\ \mu}[(\bar M_\rho^2-\partial^2)\delta^\sigma_\nu+
\partial_\nu\partial^\sigma]=\delta^\sigma_\mu.
\end{equation}
One obtains
\begin{eqnarray}\label{99.9}
{\cal L}_V&=&\frac{1}{16g^2}\ {\rm Tr}\{\partial_\mu U^\dagger
\partial^\mu U\partial_\nu U^\dagger \partial^\nu U-
\partial_\mu U^\dagger\partial_\nu U\partial^\mu U^\dagger \partial^\nu
U\}\nonumber\\ &=&\frac{1}{32g^2}[6\ {\rm Tr}\{\partial_\mu
U^\dagger\partial^\mu U\partial_\nu U^\dagger \partial^\nu U\} -({\rm
Tr}\{\partial_\mu U^\dagger\partial^\mu U\})^2\nonumber\\ &&-2{\rm
Tr}\{\partial_\mu U^\dagger\partial_\nu U\}\ {\rm Tr}\{
\partial^\mu U^\dagger\partial^\nu U\}].
\end{eqnarray}
From eq. (\ref{99.9})  one 
can infer the contribution of ${\cal L}_V$ to the parameters $L_i$
appearing in next to leading order in chiral perturbation theory
\cite{GL}
\begin{equation}\label{99.10}
L^{(V)}_1= \frac{1}{32g^2}\quad,\quad
L_2^{(V)}=\frac{1}{16g^2}\quad,\quad L_3^{(V)}=-
\frac{3}{16g^2}.
\end{equation}
We observe that the constants depend only on the effective gauge
coupling $g$.  For $g=6$ the values $L_1^{(V)}=0.87\cdot10^{-3},\
L_2^{(V)}=1.74\cdot 10^{-3},\  L_3^{(V)}=-5.2\cdot10^{-3}$ compare
well with the values  \cite{BEG} extracted from observation
$L_1=(0.7\pm0.3)\cdot10^{-3},\ L_2=(1.7\pm0.7)\cdot 10^{-3},\
L_3=-(4.4\pm2.5)\cdot10^{-3}$.  This is consistent with the hypothesis
that these constants are dominated\footnote{Further contributions
arise from the exchange of scalars  and have been estimated
\cite{JWPT} as $L_2^{(S)}=0,\ L_3^{(S)}=1.3\cdot10^{-3}$.} 
by vector-meson exchange \cite{Ecker}. The momentum  dependence of the
effective four-pion vertex extracted from inserting eq. (\ref{99.3})
 describes the
fact that the $\pi-\pi$ scattering at  intermediate energies is dominated
by the $\rho$-resonance.  We conclude that the simple effective action
(\ref{2.6}) gives a  very satisfactory picture of the pion
interactions even beyond lowest order in chiral perturbation theory!

\section{Conclusions and Discussion}
\setcounter{equation}{0}

The ``spontaneous breaking'' of color seems to be compatible
with observation. The simple effective action (\ref{2.6}) gives  a
realistic approximate description of the masses of all low-lying
mesons and baryons and of their interactions. Gluon-meson duality is
associated to the well-known Higgs phenomenon with colored composite
scalar fields,  corresponding to quark-antiquark pairs.

The most important characteristics of our scenario can be summarized
in the following points.

(i) The Higgs mechanism generates a mass for the gluons. In the limit
of equal (current) masses for the three light quarks the masses of all
gluons are equal. The massive gluons transform as an octet with
respect to the physical SU(3)-symmetry. They carry integer electric
charge, isospin and strangeness and can be identified with the vector
meson octet $\rho,K^*,\omega$.

(ii) The physical fermion fields are massive baryons with integer
electric charge. The low mass baryons form an octet with respect to
the $SU(3)$-symmetry group of the ``eightfold way''. In a
gauge-invariant language the baryons are quarks with a dressing of
nonlinear fields.  In a gauge-fixed version quarks and baryons can be
described by the same field. This is quark-baryon duality. The main
contribution to the mass of these baryons arises from chiral symmetry
breaking through quark-antiquark condensates in the color singlet and
octet channels. These considerations extend \cite{CWSSB} to the 
heavy quarks $c,b,t$, except that the mass is now dominated by 
the current quark mass.
The lightest charmed baryons (and $t$-baryons) belong to a
$SU(3)$-antitriplet with electric charge 0,1,1.  Correspondingly, the
$SU(3)$-antitriplet of the lightest $b$-baryons carries electric
charge -1,0,0.

(iii) Our scenario shares the properties of chiral symmetry breaking
with many other approaches to long-distance strong interactions. In
particular, chiral perturbation theory is recovered in the low energy
limit. This guarantees the observed mass pattern for pions, kaons and
the $\eta$-meson and the structure of their (low momentum)
interactions.

(iv) Spontaneous color symmetry breaking generates a nonlinear  local
$SU(3)_P$-reparameterization symmetry. The gauge bosons of  this
symmetry originate from the gluons. They form the octet of low masses
physical vector mesons.  The symmetry relations following from
$SU(3)_P$ symmetry appear in the electromagnetic and strong
interactions of the vector mesons and lead to vector dominance. 
They are compatible with
observation and provide for a successful test of our scenario.

(v) Weak interactions can be incorporated naturally in our framework
\cite{CWSSB}. The $\Delta I=1/2$ rule for the weak hadronic kaon decays 
turns out to be 
 a consequence of the properties of the lowest dimension operators
which are consistent with the symmetries.

Beyond the important general symmetry relations arising from chiral
symmetry and the
nonlinear local reparameterization symmetry the simple effective
action (\ref{2.6}) leads to particular predictions. They are related
to the assumption that the effective action can be described in
leading order by effective couplings with positive or zero mass
dimension. (This counting holds if composite scalar fields are
counted according to the
canonical dimension for scalar fields.) For this purpose we
fix the parameters $\chi_0, \sigma_0$ and $g$ by $M_\rho,f_\pi$ and
$\Gamma(\rho
\to e^+e^-$). More precisely, we use here eqs. (5.16), (5.17)
(7.14) with $\overline M_\rho$= 850 MeV, $f_0$= 128 MeV, $g_{\rho
\gamma}=0.12\ {\rm GeV}^2$. In
addition, the Yukawa couplings $h, \tilde h$ are fixed by the baryon
masses $M_8, M_1$ (cf. eq. (4.10)) whereas the strength of the chiral
anomaly $\nu$ and $\nu'$ determines  $M_{\eta'}^2$ by eq. (5.19). (The
precise ratio $\nu'/\nu$ is not relevant here.) For equal quark masses
we concentrate on the
following  predictions of eq.~(\ref{2.6}) which are not
dictated by symmetry considerations:

(1) The pion nucleon couplings are found as $F=D=0.5$ to be compared
with the observed values $F=0.459\pm0.008,\ D=0.798\pm 0.008$.

(2) The decay width $\Gamma(\rho\to2\pi)=115$ MeV turns out to be
somewhat lower than the observed value of 150 MeV. (Note that we use
here directly eq. (\ref{7.2}) and $\kappa_f$ equals one in leading
order, such that $g_{\rho\pi\pi}=4.6$.)

(3) The direct coupling of the photon to pions is suppressed
$g_{\gamma\pi\pi}/e=0.04$ (by virtue of eq. (\ref{E10}) with
$M_\rho\to \overline M_\rho,\ f_\pi\to f_0$).  This phenomenon of
vector dominance describes well the  observations.

(4) The effective next-to leading order couplings $L_1, L_2, L_3$ of
chiral perturbation theory come out compatible with observation
$L_2=1.7\cdot 10^{-3}, L_3=-(3.9-5.2)\cdot10^{-3}$.

For a first approximation we consider the effective action (\ref{2.6})
as very satisfactory. Corrections arise from two sources: (a)  The
nonequal quark masses break
the physical global  $SU(3)$-symmetry. (b) Higher order invariants are
certainly present in the full effective action. Some of
the corrections are known phenomenologically as, for example, the
partial Higgs effect which reduces the average meson decay constant
from $f_0$ to the mean value $f=(f_\pi+2f_K)/3$=106 MeV.  Together
with the further lowering of $f_\pi$ as compared to $f$ by $SU(3)_V$
violation this leads to a realistic  value of
$\Gamma(\rho\to2\pi)$. The neglected higher-order invariants can be
used to improve the agreement with observation. Typically, these
corrections are below 30 \%.

Many more observable quantities could be computed from the effective
action (\ref{2.6}), especially if we take into account the effects of
$SU(3)$-violation due to the strange quark mass \cite{CWMT}. (This
will involve additional parameters in the effective potential beyond
$\sigma_0$ and $\chi_0$.) We will rather give here an outlook on some
key issues for future theoretical developments. At the end, the
parameters of the effective action (\ref{2.6}) should be computed from the
microscopic action of QCD. Their values can only depend on the strong
gauge coupling (or $\Lambda_{QCD}$) and the quark masses. A first
attempt based on an instanton computation \cite{CWINS} gives a rather
satisfactory result for the value of the gluon/vector meson mass
$M_\rho$ and the associated value of the octet condensate $\chi_0$. It
also yields reasonable values for the $<\bar qq>$-condensate in the
singlet channel and for the mass and two-photon-decay width of the
$\eta'$-meson.

An interesting question concerns a variation of the quark masses. For
a large value of the strange quark mass $m_s$ (and small up and
down quark mass) an interesting Higgs picture of QCD for $N_f=2$
has been proposed in \cite{BW}. If all quark masses get heavy there
should be a transition to gluodynamics ($N_f=0$) which presumably
is characterized by an effective infrared cutoff different from the
octet condensate relevant for $N_f=3$. This raises interesting
issues for the behavior of the heavy quark potential as the quark
mass is varied \cite{CWMT} which can perhaps be tested by lattice
simulations. Other possible lattice tests concern the equivalence of
the Higgs and confinement pictures. One may introduce ``fundamental''
scalar octets with the transformation of $\chi$ together with a
classical potential. This would allow one to move continuously from the
Higgs phase with a value of $\chi_0$ independent of QCD to the QCD
universality class which can be realized in this setting by a 
 large positive mass term in
the classical potential for $\chi$. For this purpose one could vary
the octet mass term from large negative to large positive values. A
smooth transition form the Higgs phase to the QCD-universality class
would be a strong argument in favor of the possibility of a Higgs
description for the vacuum of real QCD.

Finally, our approach has all ingredients for an analytical
investigation of the QCD-phase diagram. One and the same model can
describe both quarks and gluons at high temperature $T$ and hadrons
for low $T$. The crucial qualitative change at the high temperature
phase transition (for small baryon density) is the formation of the
octet condensate as the temperature is lowered below the critical
temperature $T_c$. For $T>T_c$ one has $\chi_0=0$ and $\sigma_0$ takes
a small value which vanishes in the chiral limit of vanishing quark
masses. Both gluons and quarks are massless. In contrast, the hadronic
phase at $T<T_c$ is characterized by nonvanishing $\chi_0,\sigma_0$
and resembles the vacuum described in this work. Since the ``octet
melting'' for $T>T_c$ accounts simultaneously for both the quarks and
the gluons becoming massless\footnote{This issue has been related to
the idea of ``vector manifestation'' \cite{Rho}.}
 one finds a convincing explanation for
the equality of the ``chiral restoration temperature'' and the
``deconfinement temperature'' observed in lattice simulations. A first
quantitative computation of $T_c$ in our model (mean field
calculation) yields $T_c\approx170\ {\rm MeV}$ \cite{CWPT}.

A similar computation for a nonvanishing baryon chemical potential
still needs to be done. Recently a strong argument has advocated that
the critical temperature equals the chemical freezeout temperature and
therefore has been measured experimentally \cite{BSW}. An
extrapolation of this argument to the experiments at large baryon
density (AGS) suggests a very simple phase diagram for QCD: A first
order line separates the high-temperature quark gluon plasma (in our
picture $\chi_0=0,\sigma_0$ small) form the hadron gas
($\chi_0,\sigma_0$ close to vacuum values). This line coincides for
small $T$ and large baryon density $(\mu_B\approx m_n)$
 with the first order line between the
hadron gas and nuclear matter (nuclear gas and nuclear
liquid). Nuclear matter for $T=0$ is characterized by an additional
diquark condensate $<d_3d_3>$ \cite{CWMT} which spontaneously breaks
the global symmetries of baryon number and isospin while conserving
strangeness. In the dual hadronic description of our model this
diquark condensate corresponds to a neutron-neutron
condensate. Nuclear matter (or the nuclear liquid) is superfluid due
to the Goldstone boson associated with the spontaneous breaking of the
third component of isospin $I_3$. (Neglecting the up and down quark
masses and electroweak interactions one would have three Goldstone
bosons \cite{CWMT} associated with the spontaneous breaking of global
$SU(2)$-isospin symmetry.) The existence of a symmetry breaking order
parameter implies that the nuclear matter phase is separated by a true
phase transition line from the quark gluon plasma at high temperature
and high baryon density. This phase transition may be second order in
the $O(2)$-universality class. Critical fluctuations could
characterize the point where this line meets the first order
transition between the quark gluon plasma and the hadron gas. (Often
critical fluctuations are associated to an endpoint of the transition
between a nuclear liquid and a nuclear gas.) Finally, for low
temperature and baryon density substantially larger than nuclear
density one expects a further transition to a ``color-flavor-locked''
color superconductor \cite{4}. This is distinguished from nuclear
matter by (approximate) $SU(3)$ symmetry. The superfluidity in this
phase originates form the spontaneous breaking of baryon number while
$I_3$ remains unbroken. In the limit of equal quark masses
$m_s=m_d=m_u$ the separate phase of nuclear matter with broken $I_3$
disappears and the color superconducting phase connects directly to
the hadron gas, with an octet condensate present in both phases
\cite{CWSSB}.

Substantial work remains to be done before the Higgs picture of the
QCD vacuum can finally be accepted. Our hypothesis could be falsified
if it leads to a direct conflict with the confinement picture, lattice
simulations or observation. A better understanding of the nonlinear
fields $v, W_L, W_R$ relating quarks to baryons 
is needed in order to establish the connection to
the successful picture of the nonrelativistic quark model and to
address basic issues like the structure functions in high energy
hadron scattering.



\newpage

\begin{thebibliography}{122}
\bibitem{CWSSB}C. Wetterich, Phys. Rev. {\bf D64} (2001) 036003
\bibitem{H} G. 't Hooft, in: {\it Recent Developments in Gauge Theories}
(Plenum, New York, 1980), p. 135;\\ S. Dimopoulos, S. Raby,
L. Susskind, Nucl. Phys. {\bf B173} (1980) 208;\\ T. Matsumoto,
Phys. Lett. {\bf 97B} (1980) 131;\\ M. Yasu\`e, Phys. Rev. {\bf D42}
(1990) 3169
\bibitem{Banks} T. Banks, E. Rabinovici, Nucl. Phys. {\bf B160} (1979)
349;\\ E. Fradkin, S. Shenker, Phys. Rev. {\bf D19} (1979) 3682
\bibitem{ReWe} M. Reuter, C. Wetterich, Nucl. Phys. {\bf B408} (1993)
91;\\ C. Wetterich, ``Electroweak Physics and the Early Universe'',
eds.  J. Romao and F. Freire, Plenum Press (1994) 229;\\
W. Buchm\"uller, O. Philipsen, Nucl. Phys. {\bf B443} (1995) 47;\\
K. Kajantie, M. Laine, R. Rummukainen, M. Shaposhnikov,
Phys. Rev. Lett. {\bf 77} (1996) 2887;\\
for a review see B. Bergerhoff, C. Wetterich, hep-ph/9611462, in
``Current topics in astrofundamental physics'', eds. N.~Sanchez and
A.~Zichichi, p. 132, World Scientific 1997
\bibitem{SU2L} P. Damgaard, U. Heller, Phys. Lett. {\bf 171 B}
(1986), 442; Nucl. Phys. {\bf B294} (1987) 253; {\bf B304} (1988)
63;\\ H. Evertz, J. Jersak, K. Kanaya, Nucl. Phys. {\bf B285} (1987)
229
\bibitem{1} C. Wetterich, Z. Phys. {\bf C57} (1993) 451
\bibitem{4} D. Bailin, A. Love, Phys. Rep. {\bf 107} (1984) 325;\\
M. Alford, K. Rajagopal, F. Wilczek,  Phys. Lett. {\bf 422B} (1998)
247; Nucl. Phys. {\bf B537} (1999) 443;\\ R. Rapp, T. Sch\"afer,
E. Shuryak, M. Velkovsky, Phys. Rev. Lett.  {\bf 81} (1998) 53;\\
J. Berges, K. Rajagopal, Nucl. Phys. {\bf B538} (1999) 214
\bibitem{CF}T. Sch\"afer, F. Wilczek, Phys. Rev. Lett. {\bf 82} (1999) 3956;\\
M. Alford, J. Berges, K. Rajagopal, Nucl. Phys. {\bf B558} (1999) 219
\bibitem{HS} J. Hubbard, Phys. Rev. Lett. {\bf 3} (1959) 77;\\
R.  Stratonovich, Dokl. Akad. Nauk. SSR {\bf 115} (1957) 1097 
\bibitem{Meg} E. Meggiolaro, C. Wetterich, Nucl. Phys. {\bf B606}
(2001) 337
\bibitem{Gi} H. Gies, C. Wetterich, Phys. Rev. {\bf D69} (2004) 025001
\bibitem{Conv} C. Wetterich, Z. Phys. {\bf C48} (1990) 693
\bibitem{Anom} G. 't Hooft, Phys. Rev. {\bf D14} (1976) 3432;\\
M. Shifman, A. Vainshtein, V. Zakharov, Nucl. Phys. {\bf B163} (1980)
46;\\ M. Nowak, J. Verbaarschot, I. Zahed, Nucl. Phys. {\bf B324}
(1989) 1;\\ T. Sch\"afer, E. Shuryak, Rev. Mod. Phys. {\bf 70} (1998)
323
\bibitem{6} D. Jungnickel, C. Wetterich, Phys. Rev. {\bf D53} (1996) 5142;\\
J. Berges, D. Jungnickel, C. Wetterich, Phys. Rev. {\bf D59} (1999)
034010
\bibitem{JW} J. Jaeckel, C. Wetterich, Nucl. Phys. {\bf A733} (2004) 113
\bibitem{8} D. Jungnickel, C. Wetterich, Eur. Phys. J. {\bf C1} (1998)
669
\bibitem{Mass} D. Jungnickel, C. Wetterich, Phys. Lett. {\bf B389} (1996)
600
\bibitem{Bando}
M. Bando, T. Kugo, K. Yamawaki, Phys. Rep. {\bf 164} (1988) 217
\bibitem{KSFR} K. Kawarabayahi, M. Suzuki, Phys. Rev. Lett {\bf 16}
(1966) 255;\\ Riazuddin and Fayyazuddin, Phys. Rev. {\bf 147} (1961)
 1071
\bibitem{9} B. Serot, D. Walecka, Int. J. Mod. Phys. {\bf E6} (1997)
515
\bibitem{GL} J. Gasser, H. Leutwyler, Phys. Rep. {\bf C87} (1982) 77;
Nucl. Phys. {\bf B250} (1985) 465
\bibitem{BEG} J. Bijnens, J. G. Ecker, J. Gasser,
DAPHNE physics handbook, hep-ph/9411232
\bibitem{JWPT} D. Jungnickel, C. Wetterich, Eur. Phys. J. {\bf C2}
(1998) 557
\bibitem{Ecker} J. G. Ecker, J. Gasser, A. Pich, E. de Rafael,
Nucl. Phys. {\bf B321} (1989) 311
\bibitem{CWMT} C. Wetterich, Eur. Phys. J. {\bf C29} (2003) 251
\bibitem{CWINS} C. Wetterich, Phys. Lett. {\bf B525} (2002) 277
\bibitem{BW} J. Berges, C. Wetterich, Phys. Lett. {\bf B512} (2001) 85
\bibitem{Rho} M. Rho, hep-ph/0303136
\bibitem{CWPT} C. Wetterich, Phys. Rev. {\bf D66} (2002) 056003
\bibitem{BSW} P. Braun-Munzinger, J. Stachel, C. Wetterich,
Phys. Lett. {\bf B596} (2004) 105008
\end{thebibliography}
\end{document}